\documentclass[twocolumn]{aastex631}

\newcommand{\target}{K2-415}

\received{}
\revised{}
\accepted{}
\submitjournal{AJ}

\shorttitle{K2-415b}
\shortauthors{Hirano et al.}
\begin{document}

\title{An Earth-sized Planet around an M5 Dwarf Star at 22 pc}

\correspondingauthor{Teruyuki Hirano}
\email{hd17156b@gmail.com}

%\author[0000-0002-0786-7307]{Greg J. Schwarz}
%\affiliation{American Astronomical Society \\
%1667 K Street NW, Suite 800 \\

\author[0000-0003-3618-7535]{Teruyuki Hirano}
%\author{Teruyuki Hirano}
\affiliation{Astrobiology Center, 2-21-1 Osawa, Mitaka, Tokyo 181-8588, Japan}
\affiliation{National Astronomical Observatory of Japan, 2-21-1 Osawa, Mitaka, Tokyo 181-8588, Japan}
\affiliation{Department of Astronomical Science, School of Physical Sciences, The Graduate University for Advanced Studies (SOKENDAI), 2-21-1, Osawa, Mitaka, Tokyo, 181-8588, Japan}

\author[0000-0002-8958-0683]{Fei Dai} 
\affiliation{Division of Geological and Planetary Sciences, 1200 E California Blvd, Pasadena, CA, 91125, USA}
\affiliation{Department of Astronomy, California Institute of Technology, Pasadena, CA 91125, USA}
\affiliation{NASA Sagan Fellow}

\author[0000-0002-4881-3620]{John H. Livingston}
\affiliation{Astrobiology Center, 2-21-1 Osawa, Mitaka, Tokyo 181-8588, Japan}
\affiliation{National Astronomical Observatory of Japan, 2-21-1 Osawa, Mitaka, Tokyo 181-8588, Japan}
\affiliation{Department of Astronomical Science, School of Physical Sciences, The Graduate University for Advanced Studies (SOKENDAI), 2-21-1, Osawa, Mitaka, Tokyo, 181-8588, Japan}

\author{Sascha Grziwa}
\affiliation{Rheinisches Institut fuer Umweltforschung an der Universitaet zu Koeln, Aachener Strasse 209, 50931 Koeln, Germany}

\author{Kristine W.\ F.\ Lam}
\affiliation{Institute of Planetary Research, German Aerospace Center (DLR), Rutherfordstrasse 2, D-12489 Berlin, Germany}

\author[0000-0002-8607-358X]{Yui Kasagi}
\affiliation{Department of Astronomical Science, School of Physical Sciences, The Graduate University for Advanced Studies (SOKENDAI), 2-21-1, Osawa, Mitaka, Tokyo, 181-8588, Japan}
\affiliation{National Astronomical Observatory of Japan, 2-21-1 Osawa, Mitaka, Tokyo 181-8588, Japan}

\author[0000-0001-8511-2981]{Norio Narita}
\affiliation{Komaba Institute for Science, The University of Tokyo, 3-8-1 Komaba, Meguro, Tokyo 153-8902, Japan}
%\affiliation{JST, PRESTO, 3-8-1 Komaba, Meguro, Tokyo 153-8902, Japan}
\affiliation{Astrobiology Center, 2-21-1 Osawa, Mitaka, Tokyo 181-8588, Japan}
\affiliation{Instituto de Astrof\'{i}sica de Canarias (IAC), 38205 La Laguna, Tenerife, Spain}

\author[0000-0001-6309-4380]{Hiroyuki Tako Ishikawa}
\affiliation{Astrobiology Center, 2-21-1 Osawa, Mitaka, Tokyo 181-8588, Japan}
\affiliation{National Astronomical Observatory of Japan, 2-21-1 Osawa, Mitaka, Tokyo 181-8588, Japan}

\author[0000-0002-5706-3497]{Kohei Miyakawa}
\affiliation{National Astronomical Observatory of Japan, 2-21-1 Osawa, Mitaka, Tokyo 181-8588, Japan}

\author[0000-0001-9211-3691]{Luisa M.\ Serrano}
\affiliation{Dipartimento di Fisica, Universit\'a di Torino, Via P. Giuria 1, I-10125, Torino, Italy}

\author[0000-0002-2383-1216]{Yuji Matsumoto}
\affiliation{National Astronomical Observatory of Japan, 2-21-1 Osawa, Mitaka, Tokyo 181-8588, Japan}

\author[0000-0002-5486-7828]{Eiichiro Kokubo}
\affiliation{National Astronomical Observatory of Japan, 2-21-1 Osawa, Mitaka, Tokyo 181-8588, Japan}

\author[0000-0001-8477-2523]{Tadahiro Kimura}
\affiliation{Department of Earth and Planetary Science, Graduate School of Science, The University of Tokyo, 7-3-1 Hongo, Bunkyo-ku, Tokyo 113-0033, Japan}

\author[0000-0002-5658-5971]{Masahiro Ikoma}
\affiliation{National Astronomical Observatory of Japan, 2-21-1 Osawa, Mitaka, Tokyo 181-8588, Japan}
\affiliation{Department of Earth and Planetary Science, Graduate School of Science, The University of Tokyo, 7-3-1 Hongo, Bunkyo-ku, Tokyo 113-0033, Japan}

\author[0000-0002-4265-047X]{Joshua N.\ Winn}
\affiliation{Department of Astrophysical Sciences, Princeton University, 4 Ivy Lane, Princeton, NJ 08544, USA}

\author[0000-0001-9209-1808]{John P.\ Wisniewski}
\affiliation{George Mason University  Department of Physics \& Astronomy, 4400 University Drive, MS 3F3, Fairfax, VA 22030, USA.}

\author[0000-0002-7972-0216]{Hiroki Harakawa}
\affiliation{Subaru Telescope, 650 N. Aohoku Place, Hilo, HI 96720, USA}

\author[0000-0003-3860-6297]{Huan-Yu Teng}
\affiliation{Department of Earth and Planetary Sciences, Tokyo Institute of Technology, Meguro-ku, Tokyo, 152-8551, Japan}

\author[0000-0001-9662-3496]{William D.\ Cochran}
\affiliation{Center for Planetary Systems Habitability and McDonald Observatory, The University of Texas, Austin Texas 78730, USA}

\author[0000-0002-4909-5763]{Akihiko Fukui}
%\affiliation{Department of Earth and Planetary Science, Graduate School of Science, The University of Tokyo, 7-3-1 Hongo, Bunkyo-ku, Tokyo 113-0033, Japan}
%\affiliation{Instituto de Astrof\'{i}sica de Canarias (IAC), 38205 La Laguna, Tenerife, Spain}
\affiliation{Komaba Institute for Science, The University of Tokyo, 3-8-1 Komaba, Meguro, Tokyo 153-8902, Japan}
%\affiliation{Instituto de Astrof\'{i}sica de Canarias, V\'{i}a Lactea s/n, E-38205 La L\'{a}guna, Tenerife, Spain}
\affiliation{Instituto de Astrof\'{i}sica de Canarias (IAC), 38205 La Laguna, Tenerife, Spain}

\author[0000-0001-8627-9628]{Davide Gandolfi}
\affiliation{Dipartimento di Fisica, Universit\'a di Torino, Via P. Giuria 1, I-10125, Torino, Italy}

\author{Eike W.\ Guenther}
\affiliation{Th\"uringer Landessternwarte Tautenburg, Sternwarte 5, 07778 Tautenburg, Germany}

\author[0000-0003-4676-0251]{Yasunori Hori}
\affiliation{Astrobiology Center, 2-21-1 Osawa, Mitaka, Tokyo 181-8588, Japan}
\affiliation{National Astronomical Observatory of Japan, 2-21-1 Osawa, Mitaka, Tokyo 181-8588, Japan}
\affiliation{Department of Astronomical Science, School of Physical Sciences, The Graduate University for Advanced Studies (SOKENDAI), 2-21-1, Osawa, Mitaka, Tokyo, 181-8588, Japan}

\author[0000-0002-5978-057X]{Kai Ikuta}
\affiliation{Department of Multi-Disciplinary Sciences, Graduate School of Arts and Sciences, The University of Tokyo, 3-8-1 Komaba, Meguro, Tokyo 153-8902, Japan}

\author[0000-0003-1205-5108]{Kiyoe Kawauchi}
\affiliation{Department of Multi-Disciplinary Sciences, Graduate School of Arts and Sciences, The University of Tokyo, 3-8-1 Komaba, Meguro, Tokyo 153-8902, Japan}

\author[0000-0001-7880-594X]{Emil Knudstrup}
\affiliation{Stellar Astrophysics Centre, Department of Physics and Astronomy, Aarhus University, Ny Munkegade 120, DK-8000 Aarhus C, Denmark}

\author[0000-0002-0076-6239]{Judith Korth}
\affiliation{Department of Space, Earth and Environment, Astronomy and Plasma Physics, Chalmers University of Technology, 412 96 Gothenburg, Sweden}

\author[0000-0001-6181-3142]{Takayuki Kotani}
\affiliation{Astrobiology Center, 2-21-1 Osawa, Mitaka, Tokyo 181-8588, Japan}
\affiliation{National Astronomical Observatory of Japan, 2-21-1 Osawa, Mitaka, Tokyo 181-8588, Japan}
\affiliation{Department of Astronomical Science, School of Physical Sciences, The Graduate University for Advanced Studies (SOKENDAI), 2-21-1, Osawa, Mitaka, Tokyo, 181-8588, Japan}

\author[0000-0003-2310-9415]{Vigneshwaran Krishnamurthy}
\affiliation{Astrobiology Center, 2-21-1 Osawa, Mitaka, Tokyo 181-8588, Japan}
\affiliation{National Astronomical Observatory of Japan, 2-21-1 Osawa, Mitaka, Tokyo 181-8588, Japan}

\author[0000-0002-9294-1793]{Tomoyuki Kudo}
\affiliation{Subaru Telescope, 650 N. Aohoku Place, Hilo, HI 96720, USA}

\author{Takashi Kurokawa}
\affiliation{Astrobiology Center, 2-21-1 Osawa, Mitaka, Tokyo 181-8588, Japan}
\affiliation{Institute of Engineering, Tokyo University of Agriculture and Technology, 2-24-16, Nakacho, Koganei, Tokyo, 184-8588, Japan}

\author[0000-0002-4677-9182]{Masayuki Kuzuhara}
\affiliation{Astrobiology Center, 2-21-1 Osawa, Mitaka, Tokyo 181-8588, Japan}
\affiliation{National Astronomical Observatory of Japan, 2-21-1 Osawa, Mitaka, Tokyo 181-8588, Japan}

\author[0000-0002-4671-2957]{Rafael Luque}
\affiliation{Department of Astronomy \& Astrophysics, University of Chicago, Chicago, IL 60637, USA}

\author[0000-0003-1368-6593]{Mayuko Mori}
\affiliation{Department of Astronomy, Graduate School of Science, The University of Tokyo, 7-3-1 Hongo, Bunkyo-ku, Tokyo 113-0033, Japan}

\author[0000-0001-9326-8134]{Jun Nishikawa}
\affiliation{National Astronomical Observatory of Japan, 2-21-1 Osawa, Mitaka, Tokyo 181-8588, Japan}
\affiliation{Department of Astronomical Science, School of Physical Sciences, The Graduate University for Advanced Studies (SOKENDAI), 2-21-1, Osawa, Mitaka, Tokyo, 181-8588, Japan}
\affiliation{Astrobiology Center, 2-21-1 Osawa, Mitaka, Tokyo 181-8588, Japan}

\author[0000-0002-5051-6027]{Masashi Omiya}
\affiliation{Astrobiology Center, 2-21-1 Osawa, Mitaka, Tokyo 181-8588, Japan}
\affiliation{National Astronomical Observatory of Japan, 2-21-1 Osawa, Mitaka, Tokyo 181-8588, Japan}

\author{Jaume Orell-Miquel}
\affiliation{Instituto de Astrof\'{\i}sica de Canarias (IAC), 38205 La Laguna, Tenerife, Spain}
\affiliation{Departamento de Astrof\'\i sica, Universidad de La Laguna (ULL), 38206 La Laguna, Tenerife, Spain}

\author[0000-0003-0987-1593]{Enric Palle}
\affiliation{Instituto de Astrof\'{\i}sica de Canarias (IAC), 38205 La Laguna, Tenerife, Spain}
\affiliation{Departamento de Astrof\'\i sica, Universidad de La Laguna (ULL), 38206 La Laguna, Tenerife, Spain}

\author[0000-0003-1257-5146]{Carina M.\ Persson}
\affiliation{Department of Space, Earth and Environment, Chalmers University of Technology, Onsala Space Observatory, SE-439 92 Onsala, Sweden}

\author[0000-0003-3786-3486]{Seth Redfield}
\affiliation{Astronomy Department and Van Vleck Observatory, Wesleyan University, Middletown, CT 06459, USA}

\author{Eugene Serabyn}
\affiliation{Jet Propulsion Laboratory, California Institute of Technology, Pasadena, CA 91109, USA}

\author{Alexis M.\ S.\ Smith}
\affiliation{Institute of Planetary Research, German Aerospace Center, Rutherfordstrasse 2, 12489 Berlin, Germany}

\author{Aoi Takahashi}
\affiliation{Astrobiology Center, 2-21-1 Osawa, Mitaka, Tokyo 181-8588, Japan}
\affiliation{National Astronomical Observatory of Japan, 2-21-1 Osawa, Mitaka, Tokyo 181-8588, Japan}

\author{Takuya Takarada}
\affiliation{Astrobiology Center, 2-21-1 Osawa, Mitaka, Tokyo 181-8588, Japan}
\affiliation{National Astronomical Observatory of Japan, 2-21-1 Osawa, Mitaka, Tokyo 181-8588, Japan}

\author{Akitoshi Ueda}
\affiliation{National Astronomical Observatory of Japan, 2-21-1 Osawa, Mitaka, Tokyo 181-8588, Japan}

\author[0000-0001-5542-8870]{Vincent Van Eylen}
\affiliation{Mullard Space Science Laboratory, University College London, Holmbury St Mary, Dorking, Surrey RH5 6NT, UK}

\author[0000-0003-4018-2569]{S\'ebastien Vievard}
\affiliation{Subaru Telescope, 650 N. Aohoku Place, Hilo, HI 96720, USA}
\affiliation{Astrobiology Center, 2-21-1 Osawa, Mitaka, Tokyo 181-8588, Japan}

\author[0000-0002-6510-0681]{Motohide Tamura}
\affiliation{Department of Astronomy, Graduate School of Science, The University of Tokyo, 7-3-1 Hongo, Bunkyo-ku, Tokyo 113-0033, Japan}
\affiliation{Astrobiology Center, 2-21-1 Osawa, Mitaka, Tokyo 181-8588, Japan}
\affiliation{National Astronomical Observatory of Japan, 2-21-1 Osawa, Mitaka, Tokyo 181-8588, Japan}

\author{Bun'ei Sato}
\affiliation{Department of Earth and Planetary Sciences, Tokyo Institute of Technology, Meguro-ku, Tokyo, 152-8551, Japan}

%\author{TBD}




\begin{abstract}
We report on the discovery of an Earth-sized transiting planet ($R_p=1.015\pm0.051\,R_\oplus$)
in a $P=4.02$ day orbit around \target\ (EPIC~211414619), an M5V star at 22 pc. 
The planet candidate was first identified by analyzing the light curve data by the K2 mission, and is here shown to exist in the most recent data from TESS.
Combining the light curves with the data secured by our follow-up observations including high-resolution imaging and near infrared spectroscopy with IRD, we rule out false positive scenarios, finding a low false positive probability of $2\times 10^{-4}$. Based on IRD's radial velocities of \target, which were sparsely taken over three years, we obtain 
the planet mass of $3.0\pm 2.7\,M_\oplus$ ($M_p<7.5\,M_\oplus$ at $95\,\%$ confidence)
for \target b. 
Being one of the lowest mass stars ($\approx 0.16\,M_\odot$) known to host an Earth-sized transiting planet, \target\ will be an interesting target for further follow-up observations, including additional radial velocity monitoring and transit spectroscopy. 
\end{abstract}

\keywords{Transit photometry (1709) ---
High angular resolution (2167) ---
High resolution spectroscopy (2096) --- 
Exoplanet formation (492) --- Radial velocity (1332)}


%%%%%%%%%%%%%%%%%%%%%%%%%%
\section{Introduction} \label{sec:intro}

In the era of characterizing temperate ``Earth-like" planets from space 
\citep[e.g., with JWST;][]{2006SSRv..123..485G}
%(e.g., with JWST \citep{2006SSRv..123..485G}) 
and the ground (e.g., with 30-m class telescopes), the lowest mass stars ($\lesssim 0.3\,M_\odot$) 
are some of the highest-priority targets
for detailed atmospheric characterizations such as by transmission spectroscopy, 
by virtue of the enhanced relative scale heights 
and relative closeness of the habitable zone to the host stars 
(which provides more opportunities to observe its transits). 
Obviously, the nearest examples of low-mass stars are
of particular importance given the brightness and thus the relative ease of follow-up
observations. However, the number of known Earth-sized ($<1.25\,R_\oplus$) transiting planets around such nearby low-mass stars is very limited at this point: 14 planets in 8 systems within 30 pc as of 2023 January\footnote{\url{https://exoplanetarchive.ipac.caltech.edu}}, including the seven around TRAPPIST-1 \citep{2017Natur.542..456G}. 
%Regardless of the true habitability of those planets around M dwarfs, which are known to exhibit strong flare activity \citep{2022AN....34310079B}, they are a good laboratory to explore the conditions at which a habitable planet can exist.

The atmospheres and habitability of small rocky planets around 
pre-main-sequence \citep[pre-MS,][]{2014ApJ...797L..25R} and MS M dwarfs \citep[e.g.][]{2016ApJ...819...84K} are %have been 
a subject of debate; low-mass stars are known to be more X-ray and extreme UV (XUV) active and exhibit a higher level of flare activity \citep{2022AN....34310079B}, which can transform (or potentially blow off) a small planet's primordial atmosphere \citep[e.g.,][]{2014ApJ...792....1L} or drive the escape of water on billion-year timescales \citep[e.g.,][]{2015AsBio..15..119L, 2020ApJ...890...79J}. In addition, they spend a longer time in the pre-MS phase, implying that close-in planets around low-mass stars are more susceptible to non-thermal atmospheric erosion by high energy protons and electrons associated with coronal mass ejections and/or stellar winds \citep[e.g.,][]{2007AsBio...7..185L}. 
This atmospheric loss/evolution may also lead to the formation of dense secondary atmospheres on these small planets through geological processes such as volcanic eruptions, impact-induced outgassing \citep[e.g.][]{2008ApJ...685.1237E}, and interactions between primordial atmospheres and magma oceans \citep[e.g.][]{2006ApJ...648..696I,2021ApJ...909L..22K,2021JGRE..12606711L,2022PSJ.....3..127S}.
Thus, small planets around M dwarfs are a good laboratory to explore the atmospheric diversity of rocky planets and the conditions at which a habitable terrestrial planet can exist.

Besides the astrobiological advantages, low-mass stars are intriguing targets as regards planet formation and evolution. 
Planet demographics for low-mass stars revealed interesting patterns such as 
the population of ultra-short-period (USP) planets \citep{2018NewAR..83...37W}, 
properties of the radius and density gaps \citep[e.g.,][]{2021MNRAS.507.2154V, 2022Sci...377.1211L},
and stellar-metallicity correlation with planet size \citep[e.g.,][]{2018AJ....155..127H}, 
most of which seem to be attributable 
to the different properties of protoplanetary disks and star-planet interactions around low-mass stars \citep[e.g.,][]{2020ApJ...905...71M, 2013ApJ...775..105O, 2017ApJ...847...29O, 2015MNRAS.453.1471D}. 
There has been good evidence that the occurrence rate of close-in planets 
grows as their host stars become cooler 
\citep[e.g.,][]{2015ApJ...807...45D, 2016MNRAS.457.2877G, 2019AJ....158...75H, 2021A&A...653A.114S}, 
but the latest statistics based on the planet yields by the Transiting Exoplanet Survey Satellite \citep[TESS;][]{2015JATIS...1a4003R} suggests %that there seems to be 
a lack of detected planets around the lowest mass stars %($<0.3\,M_\odot$) 
\citep{2022AJ....163..255B}, which is at least partly supported by recent population synthesis models \citep[e.g.,][]{2021A&A...656A..72B}. 
Although these findings potentially have a significant impact on planet formation theory 
for low-mass stars, they should be corroborated with a much larger sample size, 
together with the extension towards even lower masses ($\lesssim 0.2\,M_\odot$). 
%given the number of planets around lowest-mass M dwarfs ($<0.3\,M_\odot$) is still limited. 

In this paper, we report on the detection and follow-up observations of a new transiting planet
around \target\ (EPIC~211414619), an M5V star $\approx 22$ pc from Earth (Table \ref{hyo1}). 
%\citep[Table \ref{hyo1};][]{2005AJ....129.1483L, 2006AJ....131.1163S, 2021PASJ...73..154K, 2021A&A...649A...1G}. 
%with the mass of $\approx 0.16\,M_\star$. 
{\it Kepler}'s repurposed mission, K2 \citep{Howell2014}, first obtained light curves for the star in 2017, and our analysis 
using our own transit detection/vetting pipeline identified an Earth-sized planet candidate
in a $P=4.02$ day orbit, in the framework of the KESPRINT consortium:
KESPRINT is an international consortium of scientists attempting to detect and characterize transiting exoplanets identified by space-based missions
\citep[e.g.,][]{2015ApJ...812..112S, 2017A&A...604A..16F, 2017AJ....154..123G}. 
Recently, TESS observed the ecliptic plane, which provided a unique opportunity to revisit 
a number of K2 targets including \target,
helping us validate new planet candidates and refine the ephemerides of known transiting planets. \target\ is one of the lowest mass stars ($\approx 0.16\,M_\odot$) observed by both K2 and TESS.

The rest of the paper is organized as follows. Section \ref{sec:photometry} presents the detection of the
candidate transiting planet based on the K2 and TESS observations, by which we passed
the candidate to further follow-up observations. These follow-up observations including high-resolution
imaging and spectroscopy will be described in Section \ref{sec:follow-up}, and the detailed analyses of the 
light curves and the spectroscopic data will be given in Section \ref{sec:ana}. 
In Section \ref{sec:discussion}, we discuss the future prospects for further follow-up observations. 
Section \ref{sec:summary} summarizes our findings on the new planet. 

%%%%%%%%%%%%%%%%%%%%%%%%%%%%%%%%%%%%%%%%%%%%%%%%%%%%%%%%%%%%%%%%%%%%%%
\begin{table}[t]
%\tabletypesize{\small}
\centering
\caption{Stellar Parameters of \target\ (EPIC~211414619)}\label{hyo1}
\begin{tabular}{lcc}
\hline\hline
Parameter & Value & Reference \\\hline
{\bf (Literature Values)} & & \\
LSPM ID & J0908+1151 & (a) \\
%2MASS ID & J09084885+1151411\\
%EPIC ID &  211414619 & \\
TIC ID & 323687123 & (b) \\
TOI Number & 5557 & (c) \\
$\alpha$ (J2000) & 09:08:48.855 & (d) \\
$\delta$ (J2000) & +11:51:41.116 & (d) \\
%$\mu_\alpha$ (mas yr$^{-1}$) & $81.643\pm 0.081$ & $37.818\pm 0.096$ \\
$\mu_\alpha\cos \delta$ (mas yr$^{-1}$) & $-458.503\pm 0.021$ & (d) \\
%$\mu_\delta$ (mas yr$^{-1}$) & $13.486\pm 0.058$ & $-86.911\pm 0.059$ \\
$\mu_\delta$ (mas yr$^{-1}$) & $192.574\pm 0.016$ & (d) \\
%parallax & $28.321 \pm 0.042$ &  $26.5561 \pm 0.0510$ \\
parallax (mas) & $45.8625 \pm 0.0196$ & (d) \\
$V$ (mag) & $15.330 \pm 0.027$ & (e) \\
$Gaia$ (mag) & $13.7957 \pm 0.0004$ & (d) \\
$TESS$ (mag) & $12.4289 \pm 0.0073$ & (b)\\
$J$ (mag) & $10.739 \pm 0.026$ & (f) \\
$H$ (mag) & $10.170 \pm 0.023$ & (f) \\
$K$ (mag) & $9.899 \pm 0.023$ & (f) \\
spectral type & M5V & (g) \\
EWH$\alpha$ ($\mathrm{\AA}$) & $-1.4600 \pm 0.0036$ & (h) \\\hline
{\bf (Derived Values)} & & \\
$d$ (pc) & $21.8043 \pm 0.0093$ & (i) \\
$T_\mathrm{eff}$ (K) & $3173 \pm 53$ & (i) \\
%$[\mathrm{M/H}]$ (dex) & & \\
$[\mathrm{Fe/H}]$ (dex) & $-0.13 \pm 0.18$ & (i) \\
$[\mathrm{Na/H}]$ (dex) & $-0.12\pm0.24$ & (i) \\
$[\mathrm{Mg/H}]$ (dex) & $-0.09\pm0.30$ & (i) \\
%$[\mathrm{Si/H}]$ (dex) & $0.77\pm0.31$ & \\
$[\mathrm{Ca/H}]$ (dex) & $-0.18\pm0.21$ & (i) \\
$[\mathrm{Ti/H}]$ (dex) & $0.21\pm0.34$ & (i) \\
$[\mathrm{Cr/H}]$ (dex) & $-0.17\pm0.16$ & (i) \\
$[\mathrm{Mn/H}]$ (dex) & $-0.10\pm0.26$ & (i) \\
$[\mathrm{Sr/H}]$ (dex) & $-0.05\pm0.29$ & (i) \\
$\log g$ (cgs) & $5.066 \pm 0.027$ & (i) \\
$M_\star$ ($M_\odot$) & $0.1635 \pm 0.0041$ & (i) \\
$R_\star$ ($R_\odot$) & $0.1965 \pm 0.0058$ & (i) \\
$\rho_\star$ (g cm$^{-3}$) & $30.3_{-2.6}^{+2.9}$ & (i) \\
$L_\star$ ($L_\odot$) & $	0.00351_{-0.00030}^{+0.00033}$ & (i) \\
systemic RV (km s$^{-1}$) & $22.5 \pm 0.1$ & (i) \\
$U$ (km s$^{-1}$) & $-53.72 \pm 0.07$ & (i) \\
$V$ (km s$^{-1}$) & $6.53 \pm 0.05$ & (i) \\
$W$ (km s$^{-1}$) & $-14.83 \pm 0.06$ & (i) \\
\hline
\end{tabular}
{\bf Notes.} References: (a) \citet{2005AJ....129.1483L}, (b) \citet{2019AJ....158..138S}, (c) \citet{2021ApJS..254...39G}, (d) \citet{2021A&A...649A...1G}, (e) \citet{2016yCat.2336....0H}, (f) \citet{2006AJ....131.1163S}, (g) \citet{2021PASJ...73..154K}, (h) \url{https://dr7.lamost.org/}, 
(i) this work. 
\end{table}
%%%%%%%%%%%%%%%%%%%%%%%%%%%%%%%%%%%%%%%%%%%%%%%%%%%%%%%%%%%%%%%%%%%%%%

%%%%%%%%%%%%%%%%%%%%%%%%%%
\section{Space Photometry and Detection of the Planet Candidate} \label{sec:photometry}

%\subsection{Imaging and Photometry} \label{sec:obs_im}

\subsection{K2 Photometry}\label{sec:ktwo}
\target\ was observed by K2 in Campaign 16 from UT 2017 December 13 to UT 2018 February 25 in the long-cadence (30-min) mode. 
We reduced the target pixel files downloaded from the Mikulski Archive for Space Telescopes (MAST) website\footnote{Some of the data presented in this paper were obtained from the MAST at the Space Telescope Science Institute. The specific observations analyzed can be accessed via 
\dataset[10.17909/T9K30X]{https://doi.org/10.17909/T9K30X} and \dataset[10.17909/79st-3m66]{https://doi.org/10.17909/79st-3m66}. }. 
Our pipeline from reducing K2 data, searching for transiting planets to vetting the planet candidates was detailed in \citet{2018AJ....155..127H}. The major challenge in reducing K2 data was to mitigate the systematic variation due to the rolling motion of the telescope along the boresight \citep{Howell2014}. We decorrelated the flux variation with the flux centroid motion using a method similar to that described by \citet{Vanderburg}. We then detrended the light curve with a cubic spline as a function of time with a width of 0.75-day to remove long-term systematics and stellar activity. 
We note that this spline-detrending was only used to empirically normalize the light curve for planet detection. A more physically motivated modeling of the light curve using Gaussian Process regression was also conducted as described in Section \ref{sec:RV}. 
A Box-least-square \citep[BLS;][]{Kovacs} search returned a strong detection with signal detection efficiency (SDE) of 12.9 at an orbital period of 4.02 day. We checked this transit signal for odd-even variation and deep secondary eclipse which are typical signs of a false positive due to eclipsing binaries. We only detected a $2.4\,\sigma$ odd even variation and a $0.9\,\sigma$ secondary eclipse (with the derived depth of $0.00014\pm0.00015$). \target\ hence passed our initial vetting and was promoted for further follow-up observations.

%%%%%%%%%%%%%%%%%%%
\begin{figure*}
\centering
\includegraphics[width=15cm]{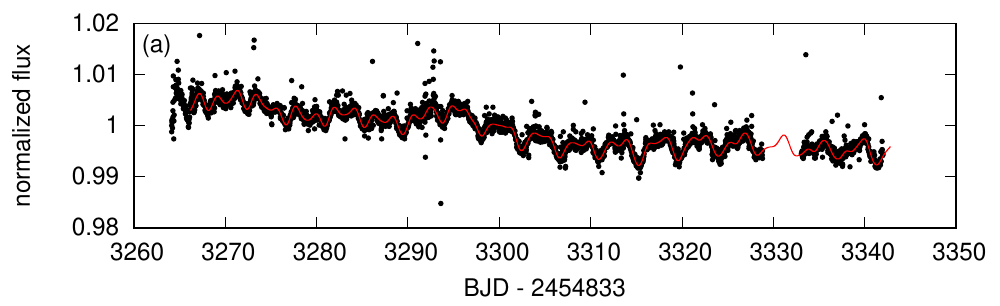}
\includegraphics[width=15cm]{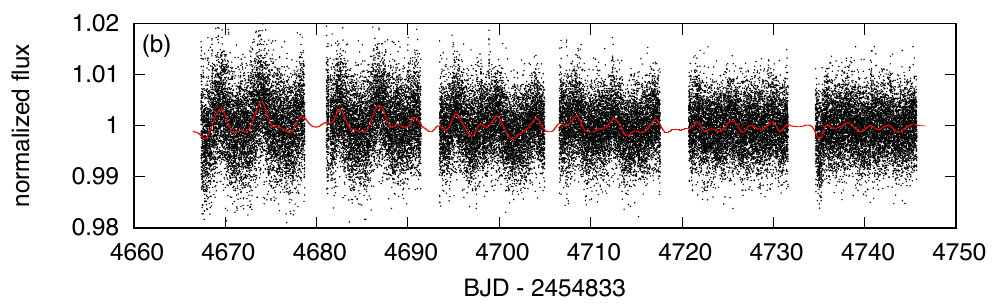}
\caption{Light curves of \target\ obtained by K2 (top; K2SFF) and TESS (bottom; PDC-SAP). 
Those data were taken at long ($\approx 29$ minutes) and short (2 minutes) cadences
for K2 and TESS light curves, respectively. The red solid line in each panel represents 
the GP regression to the observed light curve (see Section \ref{sec:RV}). 
}
\label{fig:normalizedLC}
\end{figure*}
%%%%%%%%%%%%%%%%%%%

%%%%%%%%%%%%%%%%%%%
\begin{figure}
\centering
\includegraphics[width=8.5cm]{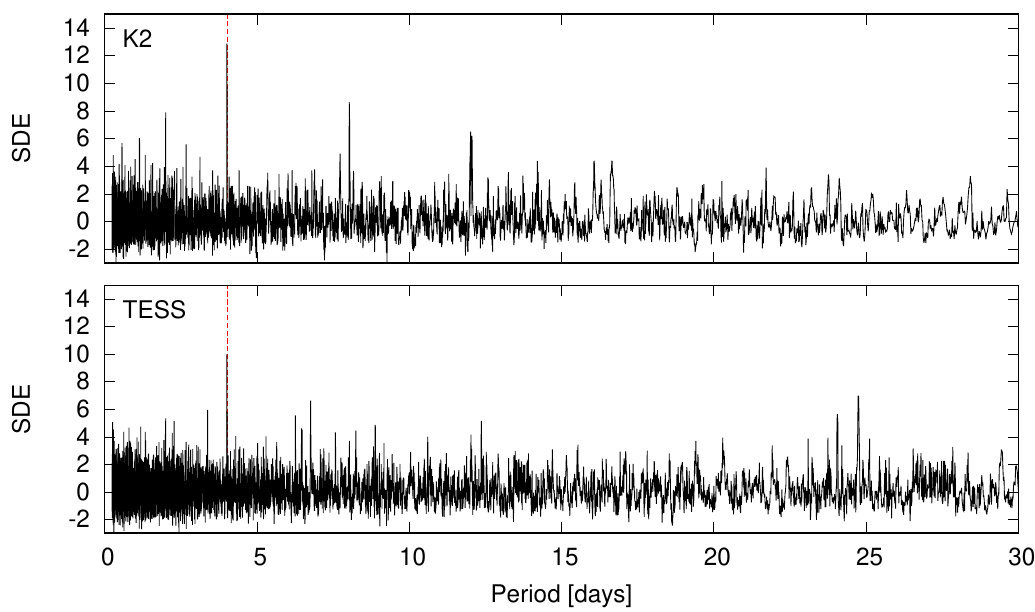}
\caption{BLS transit signal detections for the K2 (top) and TESS (bottom)
light curves. The red dashed line represents the orbital period of \target b (4.02 days). 
}
\label{fig:bls}
\end{figure}
%%%%%%%%%%%%%%%%%%%

\subsection{TESS Photometry}

\target\ was also observed by TESS at a 2-minute cadence with the TIC ID of 323687123
in Sectors 44, 45, and 46, between UT 2021 October 21 and December 30. 
In order to see if the transit signals by the same planet candidate ($P=4.02$ days)
are identifiable in the TESS data, we downloaded the PDCSAP FLUX light curve data generated by the SPOC pipeline \citep{2012PASP..124.1000S,2012PASP..124..985S,2014PASP..126..100S}. 
After applying a $3.5\,\sigma$ clipping to the flux data as well as correcting for the flux offsets 
between different sectors (panel (b) of Figure \ref{fig:normalizedLC}), 
we implemented the BLS analysis \citep{Kovacs} as in the case of K2 light curve. 
We identified the same transiting-planet candidate ($P=4.02$ days) as the highest power 
in the BLS periodogram with the SDE of 10.0 (Figure \ref{fig:bls}). 
This $P=4.02$ day planet candidate was independently detected by the TESS project, 
by which \target\ was named ``TOI-5557".

%joint analysis between K2 and TESS LCs.
In order to search for an additional transiting-planet candidate in the system, 
we performed a joint BLS analysis using both K2 and TESS light curves. 
However, no significant peak other than the $P=4.02$ day candidate was identified in the BLS periodogram.

\section{Follow-up Observations} \label{sec:follow-up}

In order to confirm the planetary nature of the transiting-planet candidate identified 
in the K2 and TESS data, we conducted follow-up observations as below. 

\subsection{AO Imaging with Subaru / IRCS}

On UT 2018 June 18, we performed high-resolution imaging for \target\ using the InfraRed Camera and Spectrograph 
\citep[IRCS;][]{2000SPIE.4008.1056K}
and the adaptive-optics system AO188 \citep{2008SPIE.7015E..10H}, both mounted on the Subaru 8.2m telescope. 
Adopting the fine-sampling mode ($1\,\mathrm{pix}=20$ mas) and the $K^\prime-$band filter ($\approx 2.1\,\mu$m),
we imaged \target\ with a five-point dithering. 
The exposure time for each dithering position was set to $3\,\mathrm{sec}\times 3\,\mathrm{coadd}=9\,\mathrm{sec}$
so that the peak count of \target's image stays within the linearity count regime. 
We obtained two sequences of the dithering pattern, giving a total on-sky integration time of 90 sec.

%%%%%%%%%%%%%%%%%%%
\begin{figure}
\centering
\includegraphics[width=8.5cm]{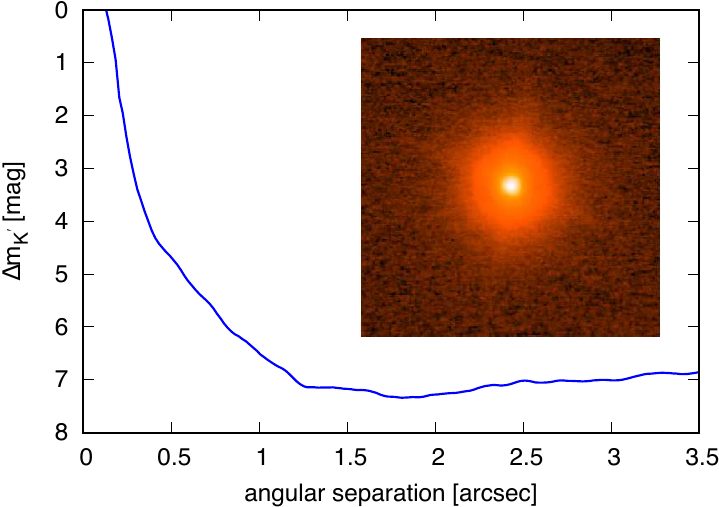}
\caption{Sensitivity plot ($5\,\sigma$ contrast curve) for \target\ in the $K^\prime$ band based on the combined IRCS image. The inset shows the zoomed image of the target with a
FoV of $4^{\prime\prime}\times4^{\prime\prime}$. 
}
\label{fig:ircs}
\end{figure}
%%%%%%%%%%%%%%%%%%%
Raw IRCS frames were reduced as in \citet{2016ApJ...820...41H}, and we aligned and median combined the reduced images 
%to achieve a high signal-to-noise (S/N) ratio. 
to suppress the background noise and thus achieve a higher flux contrast. 
The background flux scatter in the combined image was found to be 5.9 ADU
while the flux peak of \target\ was $\approx 11400$ ADU.
The full width at the half maximum (FWHM) of \target's image was measured to be 
$0\farcs133$, and no secondary source was detected in the field of view (FoV) of 
$20^{\prime\prime}\times20^{\prime\prime}$. 
To gain a constraint on the magnitude of any secondary source, we computed a $5\,\sigma$
contrast as a function of angular separation from the target, following the procedure in \citet{2016ApJ...820...41H}. 
The $5\sigma$ contrast curve is plotted in Figure \ref{fig:ircs}, in which the inset displays the $4^{\prime\prime}\times4^{\prime\prime}$
image of \target. IRCS AO imaging achieved $\Delta m_{K^\prime}$ of $6-7$ mag at the angular 
separation of $1^{\prime\prime}$.

\subsection{Speckle Observation with WIYN / NESSI}
\label{sec:speckle}

\begin{figure}
\centering
\includegraphics[width=0.48\textwidth]{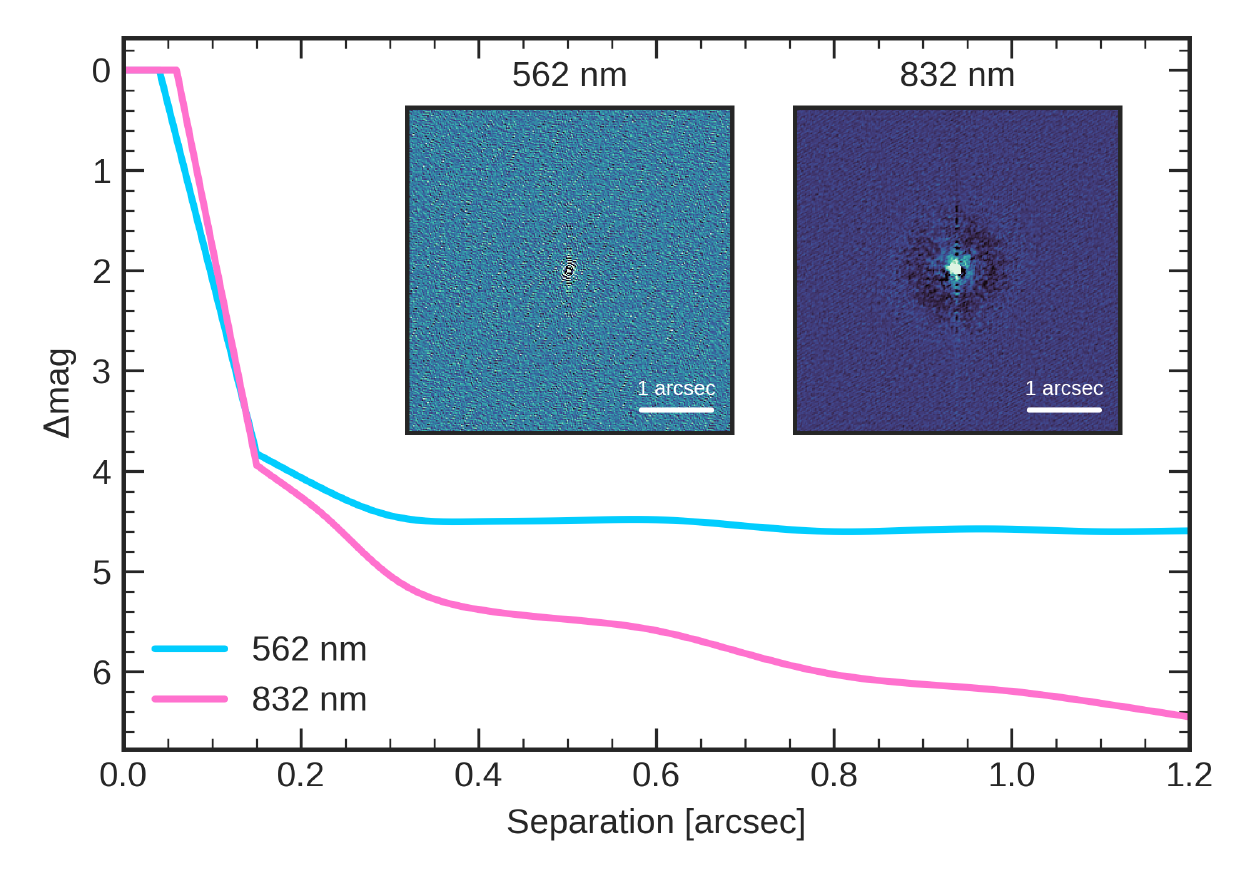}
\caption{
Sensitivity plots of \target\ based on the speckle observations by NESSI. 
The two insets display the reconstructed images of \target\ in the 562 nm (left)
and 832 nm (right) bands. 
}
\label{fig:speckle}
\end{figure}

On the night of UT 2019 January 20, \target\ was observed with the NESSI speckle imager \citep{Scott2019}, mounted on the 3.5\,m WIYN telescope at Kitt Peak. NESSI simultaneously acquires data in two bands centered at 562\,nm and 832\,nm using high speed electron-multiplying CCDs (EMCCDs). We collected and reduced the data following the procedures described in \citet{Howell2011}. The resulting reconstructed 832\,nm band image achieved a contrast of $\Delta\mathrm{mag} \sim 4$ and $\Delta\mathrm{mag} \sim 6$ at separations of 0.2\arcsec\ and 1\arcsec, respectively (see Figure~\ref{fig:speckle}). 

%%%%%%%%%%%%%%%%%%
%%%%%%%%%%%%%%%%%%

%\subsection{Spectroscopy} \label{sec:obs_sp}

\subsection{High-Resolution Spectroscopy with Subaru / IRD}\label{sec:obs-ird}

As part of the follow-up campaign for transiting planet candidates
identified by K2 and TESS, we conducted high-resolution spectroscopy using the 
InfraRed Doppler spectrograph \citep[IRD;][]{2012SPIE.8446E..1TT,2018SPIE10702E..11K} on the Subaru telescope. 
IRD covers the near IR wavelengths from 970 nm to 1730 nm with a spectral resolution of
$R\approx 70,000$. 
The IRD observations were carried out between UT 2019 January 15 and 2022 May 25, 
and a total of 42 IRD frames were obtained, including a few frames taken in the 
Subaru Strategic Program (SSP), which is a blind Doppler survey to find exoplanets 
around nearby mid-to-late M dwarfs \citep[see e.g.,][]{2022PASJ...74..904H}. 
For the nine frames secured in 2022 January, only $H-$band spectra ($1450\,\mathrm{nm}<\lambda$) 
were obtained due to the technical issues in the $YJ-$band detector. 
We set each integration time to $300-1800$ sec with a median exposure of 1200 sec. 
For each integration, we simultaneously took the reference spectrum of the laser-frequency comb 
(LFC), which is used for the estimation of the instantaneous instrumental profile (IP) of the 
spectrograph.

Raw IRD frames were reduced as in \citet{2020PASJ...72...93H}, and we extracted the 
wavelength-calibrated one-dimensional (1D) spectra for both stellar and reference (LFC) fibers. 
Based on those reduced spectra, we measured precise RVs using IRD's RV-analysis pipeline
\citep{2020PASJ...72...93H}; 
%in the RV analysis, a telluric-free, IP-deconvolved template spectrum for \target\ was
%first extracted, and the RVs for individual spectra were measured with respect to this 
%template spectrum by forward modeling. 
briefly speaking, the instantaneous IP for each spectral segment was first estimated from the 
LFC spectrum, with which individual stellar spectra were deconvolved. The stellar template
spectrum for \target\ was then generated by combining those IP-deconvolved spectra, in which
telluric lines were also removed via theoretical telluric model fits or using a rapid rotator's spectrum. The combined template was found to have good signal-to-noise (S/N) ratios (typically $>100$ per pixel) 
for the whole spectral range covered by IRD. 
Finally, relative RV ($v_\star$) with respect to this template (denoted by $S(\lambda)$) was measured by forward modeling 
each observed spectral segment ($f_\mathrm{obs}(\lambda)$) as
%%%%%%%%%%%%%%%
\begin{eqnarray}
\label{eq:pipeline}
f_\mathrm{obs} (\lambda) = k(\lambda) %\nonumber\\
\left[S\left(\lambda\sqrt{\frac{1+v_\star/c}{1-v_\star/c}}\right)T\left(\lambda\right) \right] 
*\mathrm{IP},~
\end{eqnarray}
%%%%%%%%%%%%%%%
where $T(\lambda)$ is the theoretical telluric model and $k(\lambda)$ is a second-order polynomial describing the spectrum continuum for each segment \citep[see][for more detail]{2020PASJ...72...93H}. 
The resulting relative RVs are listed in Table \ref{tab:ird}. 
The typical RV uncertainty (internal error) was $4-7$ m s$^{-1}$. 
We also estimated the absolute RV of \target\ by fitting the individual molecular
lines in the IRD spectra by Gaussian functions, and comparing their centers
with the vacuum line positions in the literature. %\citep{}. 
Based on the analysis of 37 relatively deep OH lines in the $H-$band, we obtained a mean absolute 
RV of $22.5\pm0.1$ km s$^{-1}$. 

%%%%%%%%%%%%%%%%%%%%%%%%%%%%%%%%%%%%%%%%%%%%%%%%%%%%%%%%%%%%%%%%%%%%%%
\begin{table*}[t]
%\tabletypesize{\small}
%\centering
\begin{center}
\caption{RVs and spectral indices extracted from IRD spectra}\label{tab:ird}
\begin{tabular}{lrrrrrr}
\hline\hline
Time & relative RV & RV error & $\Delta\mathrm{FWHM}$ & $\Delta\mathrm{FWHM}$ error & $\Delta\mathrm{BIS}$ & $\Delta\mathrm{BIS}$ error\\
(BJD$_\mathrm{TDB}$) & (m s$^{-1}$) & (m s$^{-1}$) & (km s$^{-1}$) & (km s$^{-1}$) & (km s$^{-1}$) & (km s$^{-1}$) \\\hline
2458498.9140269 & 6.1 & 4.7 & -0.024 & 0.038 & 0.19 & 0.18\\
2458498.9293007 & -6.5 & 5.5 & -0.170 & 0.044 & -0.10 & 0.11\\
2458499.8875579 & -15.1 & 4.6 & -0.115 & 0.036 & 0.14 & 0.18\\
2458499.9053351 & -3.5 & 4.4 & -0.126 & 0.034 & 0.20 & 0.13\\
2458499.9231437 & 3.1 & 4.4 & -0.087 & 0.033 & 0.23 & 0.14\\
2458563.8273755 & 0.8 & 3.7 & -0.066 & 0.036 & 0.06 & 0.15\\
2458563.8434206 & -0.2 & 4.1 & -0.005 & 0.041 & 0.18 & 0.14\\
2458563.8576999 & -2.6 & 4.3 & -0.113 & 0.040 & 0.02 & 0.13\\
2458626.8089605 & -3.5 & 10.9 & -0.151 & 0.102 & -0.02 & 0.21\\
2458626.8127899 & -33.4 & 10.2 & -0.351 & 0.091 & 0.10 & 0.20\\
2458626.8164678 & 0.8 & 10.2 & -0.149 & 0.095 & 0.15 & 0.24\\
2458649.7618439 & 3.1 & 6.5 & -0.229 & 0.063 & -0.03 & 0.13\\
2458649.7691465 & -3.3 & 7.1 & -0.068 & 0.069 & 0.09 & 0.18\\
2458649.7764790 & 7.5 & 7.1 & -0.078 & 0.071 & -0.07 & 0.13\\
2458828.0584522 & -0.7 & 4.9 & -0.100 & 0.047 & 0.02 & 0.14\\
2458896.0058500 & 10.7 & 10.0 & -0.109 & 0.092 & 0.43 & 0.26\\
2458896.0167066 & 0.2 & 7.2 & 0.043 & 0.068 & 0.53 & 0.26\\
2459244.9776946 & -3.7 & 4.8 & 0.014 & 0.045 & 0.14 & 0.21\\
2459247.9043374 & -3.2 & 4.9 & 0.051 & 0.045 & -0.32 & 0.25\\
2459275.9973405 & -6.2 & 6.1 & -0.020 & 0.057 & -0.08 & 0.13\\
2459276.0151846 & 16.0 & 9.8 & -0.205 & 0.089 & -0.07 & 0.20\\
2459321.8971738 & 27.1 & 5.8 & -0.124 & 0.055 & 0.06 & 0.47\\
2459321.9113889 & 4.8 & 5.4 & 0.029 & 0.050 & -0.15 & 0.15\\
2459323.7582073 & 12.9 & 4.5 & 0.143 & 0.041 & 0.10 & 0.17\\
2459323.7724223 & 16.3 & 4.5 & 0.021 & 0.042 & -0.09 & 0.15\\
2459328.7606870 & 29.7 & 5.5 & -0.020 & 0.050 & -0.03 & 0.13\\
2459328.7749031 & 12.0 & 5.3 & -0.069 & 0.052 & 0.13 & 0.21\\
2459337.7794598 & 4.0 & 9.6 & 0.040 & 0.091 & 0.14 & 0.23\\
2459338.7871978 & 2.6 & 5.0 & -0.044 & 0.040 & 0.03 & 0.11\\
2459588.1189335 & 12.8 & 6.4 & 0.055 & 0.042 & 0.41 & 0.17\\
2459589.1004307 & 14.3 & 6.1 & -0.036 & 0.038 & 0.19 & 0.18\\
2459589.1182238 & 17.9 & 6.1 & -0.163 & 0.036 & 0.22 & 0.13\\
2459596.8815868 & 2.0 & 6.4 & -0.026 & 0.049 & 0.39 & 0.23\\
2459596.8996175 & -9.9 & 6.4 & -0.008 & 0.049 & 0.55 & 0.21\\
2459602.1065831 & -2.3 & 11.3 & 0.478 & 0.097 & -0.24 & 0.35\\
2459602.1209018 & 4.1 & 12.3 & 0.347 & 0.104 & -0.04 & 0.32\\
2459604.0756268 & -8.3 & 5.2 & 0.072 & 0.040 & -0.01 & 0.21\\
2459604.0969686 & -20.9 & 5.2 & 0.029 & 0.040 & 0.01 & 0.22\\
2459648.9839188 & -10.0 & 8.9 & -0.013 & 0.081 & 0.28 & 0.18\\
2459711.7427174 & -14.1 & 6.3 & -0.035 & 0.058 & 0.32 & 0.16\\
2459711.7569324 & -12.9 & 6.3 & 0.010 & 0.060 & 0.22 & 0.18\\
2459724.7450558 & -2.9 & 4.8 & -0.045 & 0.046 & 0.22 & 0.12\\
\hline
\end{tabular}
\end{center}
\end{table*}
%%%%%%%%%%%%%%%%%%%%%%%%%%%%%%%%%%%%%%%%%%%%%%%%%%%%%%%%%%%%%%%%%%%%%%

%\subsubsection{Subaru / HDS Spectroscopy}

%%%%%%%%%%%%%%%%%%%%%%%%%%
\section{Analyses and Results} \label{sec:ana}

\subsection{Estimation of Stellar Parameters} \label{sec:parameters}

%estimation of atmospheric paramters
Using \target's template spectrum produced in Section \ref{sec:obs-ird}, we performed a 
line-by-line analysis to estimate the stellar atmospheric parameters. 
We measured equivalent widths of FeH molecular lines and atomic lines of Na, Mg, Ca, Ti, Cr, Mn, Fe, and Sr to derive an effective temperature $T_{\mathrm{eff}}$ and abundances of individual elements [X/H].
For the $T_{\mathrm{eff}}$ estimation, 47 well-isolated FeH lines in the Wing-Ford band at $990-1020$ nm were employed in the procedure described by \citet{2022AJ....163...72I}.
For the abundance analysis, we analyzed 33 atomic lines by following the procedure of \citet{2020PASJ...72..102I}.

We iterated the $T_{\mathrm{eff}}$ estimation and the abundance analysis alternately until both results became consistent with each other.
First, we derived a provisional $T_{\mathrm{eff}}$ adopting the solar metallicity as the initial value, and then by adopting this provisional $T_{\mathrm{eff}}$, we derived the individual abundances of the eight elements.
Subsequently, we adopted the iron abundance [Fe/H] derived in the previous step as the input metallicity of the $T_{\mathrm{eff}}$ estimation. 
Finally, adopting the new $T_{\mathrm{eff}}$ value, we redetermined the elemental abundances.
The procedure up to this point allows the inputs and outputs of the analyses to be consistent within the measurement errors.

The final results of the $T_{\mathrm{eff}}$ and the elemental abundances are listed in Table \ref{hyo1}.
Note that the uncertainty of the $T_{\mathrm{eff}}$ here is calculated by the standard deviation of the estimates from the individual FeH lines, while it may also have a systematic error of less than 100 K as discussed in \citet{2022AJ....163...72I}.

%estimation of other parameters
Based on those atmospheric parameters, we also derived the other physical parameters of \target. 
Inputting the above-derived metallicity [Fe/H] as well as 2MASS's $K_s-$band magnitude \citep{2006AJ....131.1163S} 
and Gaia parallax \citep{2021A&A...649A...1G}, 
we derived the stellar mass $M_\star$ and radius $R_\star$ using the empirical relations by \citet{2019ApJ...871...63M} 
and \citet{2015ApJ...804...64M}, 
respectively. The uncertainties of those parameters were calculated with Monte Carlo simulations
assuming Gaussian distributions for the above input parameters and systematic errors in the empirical
formula \citep{2019ApJ...871...63M}. 
In the Monte Carlo calculations, we derived the surface gravity $\log g$, mean stellar density $\rho_\star$, and luminosity $L_\star$. 
We also computed the Galactic $(U,V,W)$ velocities with respect to the Sun, using
the Gaia DR3 information as well as the systemic RV (Table \ref{hyo1}). 
All of those derived parameters are also summarized in Table \ref{hyo1}. 
The low kinematic velocities indicate that \target\ is a thin disk star.

%estimation of rotation period

%%%%%%%%%%%%%%%%%%%
\begin{figure}
\centering
\includegraphics[width=8.5cm]{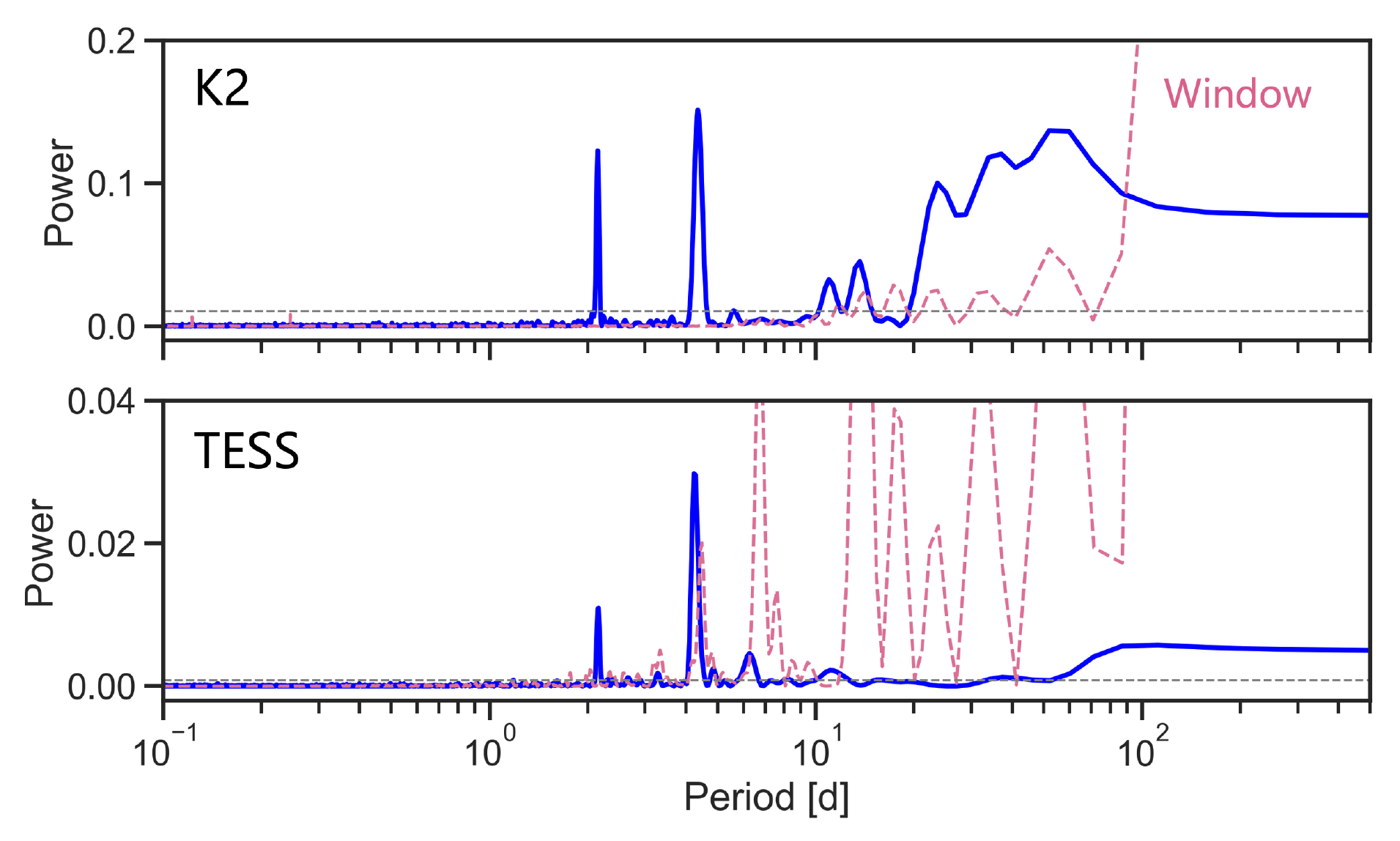}
\caption{
LS periodograms for the K2 (top) and TESS (bottom) light curves. 
The window function for each light curve is plotted by the dashed line. 
The secondary peak at $\approx 2$ days represents the harmonics of the rotation period. 
}
\label{fig:K2-TESS_LS}
\end{figure}
%%%%%%%%%%%%%%%%%%%
Both K2 and TESS light curves in Figure \ref{fig:normalizedLC} exhibit flux variations 
that are representative of spot-induced rotational modulations. 
To estimate the
rotation period of \target, we computed the generalized Lomb-Scargle (LS) periodogram
\citep{2009A&A...496..577Z} for the two light curves. 
As shown in Figure \ref{fig:K2-TESS_LS}, both periodograms show strong peaks 
at similar periods, whose powers correspond to false-alarm probabilities (FAP) 
much lower than $10^{-3}$. Inspecting the peaks of the periodograms as well as
the shapes of original light curves, we estimated the rotation period of the star 
to be $P_\mathrm{rot}=4.36 \pm 0.15$ days for K2 and $P_\mathrm{rot}=4.26\pm0.12$
days for TESS light curves, respectively. 
The uncertainties of these periods were derived based on the FWHMs of the peaks.

%%%%%%%%%%%%%%%%%%%
\begin{figure}
\centering
\includegraphics[width=8.5cm]{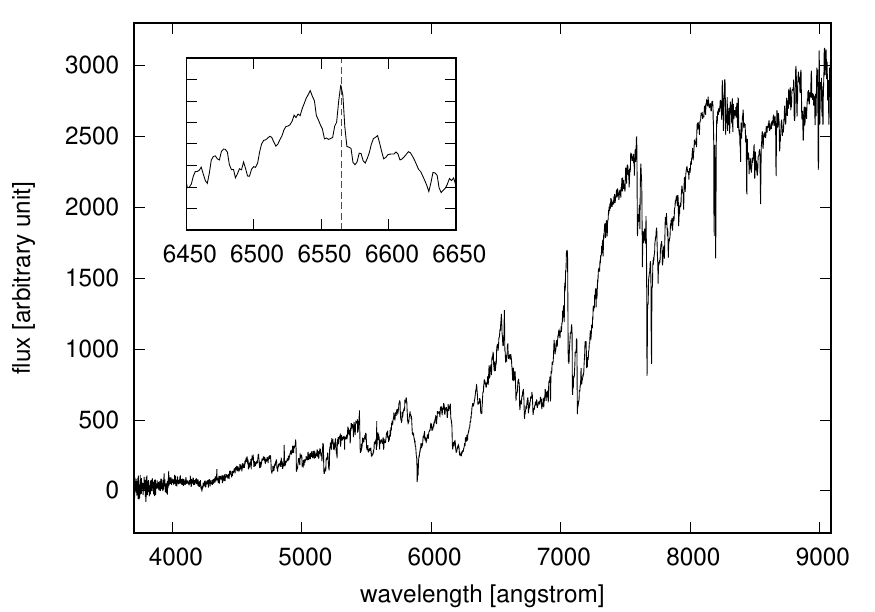}
\caption{
Low-resolution optical spectrum of \target\ obtained by LAMOST \citep{2015RAA....15.1095L}. 
The inset displays a zoom-in around H$\alpha$ (the red vertical line). 
}
\label{fig:lamost}
\end{figure}
%%%%%%%%%%%%%%%%%%%
The clear photometric variability (with the amplitude of $0.2-0.4\,\%$) as well as the relatively short period of rotation suggests that \target\ is a moderately active star. Inspecting an archived optical spectrum of \target\ observed by the Large Sky Area Multi-Object Fiber Spectroscopic Telescope \citep[LAMOST,][]{2015RAA....15.1095L}, we checked for its chromospheric activity. As shown in Figure \ref{fig:lamost}, \target's low-resolution spectrum exhibits a moderate emission at H$\alpha$, whose EW is measured to be $\mathrm{EWH}\alpha=-1.4600 \pm 0.0036\,\mathrm{\AA}$\footnote{\url{https://dr7.lamost.org}} (a negative value of EW indicates that it is an emission line), in agreement with the moderate variability seen in the light curves.
We note that the variability amplitude of light curve modulations seems to evolve rapidly in time, as evidenced by the vanishing variation in the second half of the TESS data. This fact suggests that activity-induced spectroscopic variations such as RV jitters do not produce a coherent pattern even on a timescale of $1-2$ months. We will discuss the spot-induced photometric and spectroscopic variabilities in more detail in Section \ref{sec:RV}.

\subsection{Joint Analyses of Transit Photometry} \label{sec:ana_photometry}

Since two different sets of light curves (K2 and TESS) are available for \target, we first analyzed each data set independently
to check for consistency in the derived parameters (e.g., transit depth).  
For the transit analysis of K2 data, we ended up using the public light curve, K2SFF \citep{2014PASP..126..948V}, as K2SFF generally 
delivers better-quality light curves than our own ones (Section \ref{sec:ktwo}) 
in terms of flux scatters and behavior of outliers. 
Following the procedure in \citet{2015ApJ...799....9H}, we extracted light curve segments around the transits (assuming a constant period, with a period
based on a preliminary fit), and simultaneously fitted all the extracted segments allowing for possible transit timing 
variations (TTVs). Each extracted segment covers approximately $5\times$ the transit duration, thus typically involving 
$8-10$ flux points. We implemented a Markov Chain Monte Carlo (MCMC) analysis to fit the light curve segments after a global optimization of the fitting parameters using Powell's conjugate direction method as in \citet{2015ApJ...799....9H}. 
In the analysis, 
%we employed the transit model by \citet{2009ApJ...690....1O} with an integration time of the $\approx 29$-minute cadence and a linear function of time for the out-of-transit baseline. 
we employed the transit model by \citet{2009ApJ...690....1O} and a linear function of time for the out-of-transit baseline. To take into account the $\approx 29$-minute cadence of K2 photometry, for each observed flux we computed the transit model with one-minute sampling and binned the light curve to the K2 sampling. 
The fitting parameters are the scaled semi-major axis $a/R_\star$, transit impact parameter $b$, star-to-planet radius ratio $R_p/R_\star$, limb-darkening parameters in the quadratic law ($u_1$ and $u_2$), orbital period $P$, and mid-transit time for each transit $T_c^{(i)}$. 
We fixed the orbital eccentricity at $e=0$ at this point. Only for the limb-darkening parameters, we imposed Gaussian priors as $u_1+u_2=0.81\pm0.20$ and $u_1-u_2=0.00\pm0.20$, based on the theoretical calculations by \citet{2013A&A...552A..16C}. The result of this MCMC analysis is given in the leftmost column of Table \ref{hyo2}. 
From the mid-transit times for individual transits, we also computed the zero-point mid-transit time $T_{c,0}$ for the K2 data set assuming a linear ephemeris (i.e., a constant period).

We also analyzed the TESS light curve (PDCSAP FLUX) following the same procedure as above. 
The only differences in the modeling are that we did not integrate the theoretical transit model for TESS data (2-minute cadence) 
and we adopted different priors for the limb-darkening parameters 
($u_1+u_2=0.73\pm0.20$ and $u_1-u_2=-0.17\pm0.20$ for the TESS band). 
The result of the MCMC fit to the TESS light curve is presented in the second column of Table \ref{hyo2}.

%%%%%%%%%%%%%%%%%%%
\begin{figure}
\centering
\includegraphics[width=8.5cm]{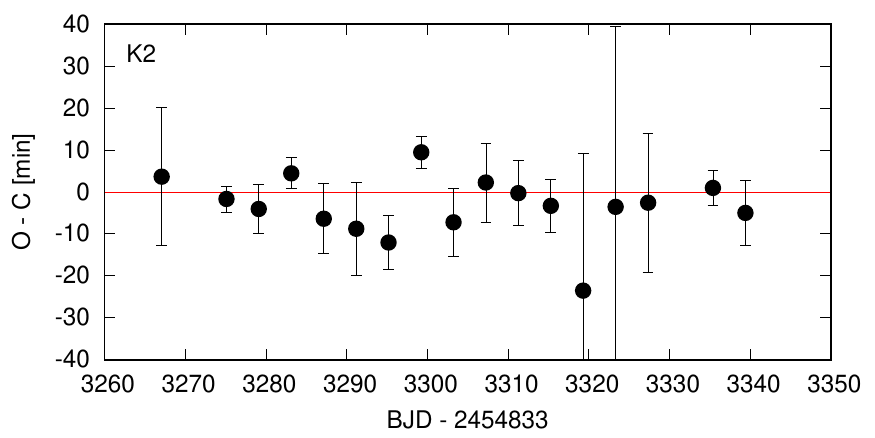}
\includegraphics[width=8.5cm]{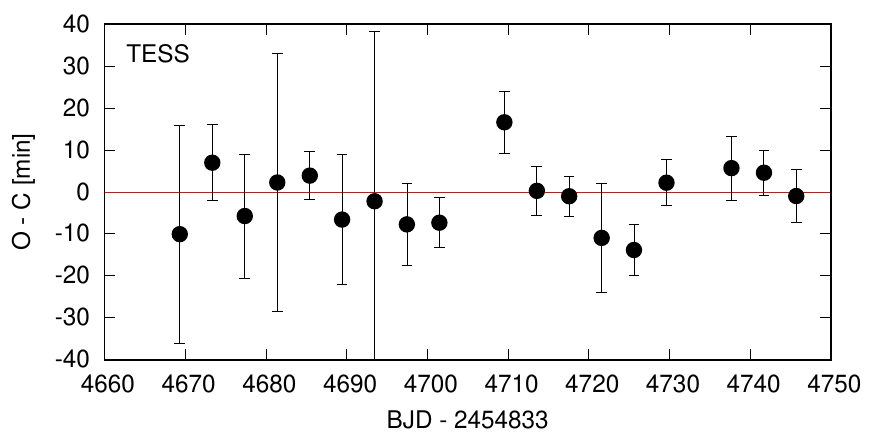}
\caption{Residuals from the linear ephemeris (a constant period) for
\target b's transit times for the K2 (top) and TESS (bottom) data. 
}
\label{fig:ttv}
\end{figure}
%%%%%%%%%%%%%%%%%%%
The transit parameters ($R_p/R_\star$ and $P$ in particular)
derived from K2 and TESS light curves are consistent with one another within $1\sigma$, 
suggesting that no significant dilution is blended in those light curves given those were
obtained with different apertures and different pass bands. 
The stellar density estimated from the transit modeling is $47_{-39}^{+51}$ g cm$^{-3}$
for K2 and $24_{-17}^{+13}$ g cm$^{-3}$ for TESS light curves, respectively. Although these are
both compatible with the mean stellar density derived in 
Section \ref{sec:parameters} ($\rho_\star = 30.3_{-2.6}^{+2.9}$ g cm$^{-3}$), 
the constraint is rather weak due to the degeneracy in $a/R_\star$, $b$, and $R_p/R_\star$ 
in the transit modeling. 
Figure \ref{fig:ttv} plots the observed minus calculated ($O-C$) residuals from the linear ephemeris for the transit times derived from the K2 (top) and TESS (bottom) data sets. 
No significant TTVs are seen for both data sets; 
the $\chi^2$ statistics for the transit times are 15.2 (K2) and 16.3 (TESS) with the degrees of freedom being 15 and 16, respectively. 
We also ensured that the two $T_{c,0}$ values derived from the K2 and TESS data sets are 
compatible with each other once $P$'s error propagation is taken into account.

%%%%%%%%%%%%%%%%%%%%%%%%%%%%%%%%%%%%%%%%%%%%%%%%%%%%%%%%%%%%%%%%%%%%%%
\begin{table*}[t]
%\tabletypesize{\small}
\centering
\caption{Results of Transit Analyses for \target b}\label{hyo2}
\begin{tabular}{lcccc}
\hline\hline
Data Set & K2 ($e=0$) & TESS ($e=0$) & K2 + TESS ($e=0$) & K2 + TESS ($e$ prior) \\\hline
(Fitting Parameter) & & & &  \\
$a/R_\star$ & $34_{-16}^{+10}$ & $27.4_{-9.3}^{+4.2}$ & $29.74_{-0.91}^{+0.83}$ & $29.63_{-0.91}^{+0.86}$  \\
$b$ & $0.52_{-0.36}^{+0.39}$ & $0.43_{-0.30}^{+0.37}$ & $0.39_{-0.18}^{+0.11}$ & $0.37_{-0.23}^{+0.19}$ \\
$R_p/R_\star$ & $0.0513_{-0.0045}^{+0.0138}$ & $0.0514_{-0.0033}^{+0.0053}$ & $0.0470\pm 0.0018$ & $0.0472_{-0.0018}^{+0.0020}$ \\
$u_1^{(\mathrm{K2})}+u_2^{(\mathrm{K2})}$ & $0.81\pm 0.20$ & $-$ & $0.79\pm 0.18$ & $0.79\pm 0.19$\\
$u_1^{(\mathrm{K2})}-u_2^{(\mathrm{K2})}$ & $-0.01\pm 0.20$ & $-$ & $0.00\pm 0.20$ & $-0.01\pm 0.20$ \\
$u_1^{(\mathrm{TESS})}+u_2^{(\mathrm{TESS})}$ & $-$ & $0.72\pm 0.20$ & $0.74 \pm 0.20$ & $0.72\pm 0.19$ \\
$u_1^{(\mathrm{TESS})}-u_2^{(\mathrm{TESS})}$ & $-$ & $-0.17\pm 0.20$ & $-0.16 \pm 0.20$ & $-0.17\pm 0.20$ \\
$\sqrt{e}\cos\omega$ & 0 (fixed) & 0 (fixed) & 0 (fixed) & $0.00_{-0.30}^{+0.29}$  \\
$\sqrt{e}\sin\omega$ & 0 (fixed) & 0 (fixed) & 0 (fixed) & $0.00_{-0.25}^{+0.20}$ \\
$P$ (days) & $4.01780 \pm 0.00019$ & $4.01803 \pm 0.00024$ & $4.0179682\pm 0.0000021$ & $4.0179694\pm 0.0000027$ \\
$T_{c,0}-2454833$ (BJD) & $3267.0727\pm0.0018$ & $4665.3230 \pm 0.0033$ & $3267.07137\pm 0.00054$ & $3267.07116\pm 0.00069$ \\
\hline
\end{tabular}
\end{table*}
%%%%%%%%%%%%%%%%%%%%%%%%%%%%%%%%%%%%%%%%%%%%%%%%%%%%%%%%%%%%%%%%%%%%%%

Satisfied with the result of consistency checks, we next performed joint analyses of K2 and TESS light curves. 
Based on the lack of TTVs, in the joint analyses we did not allow for a variable $T_c$ for each transit, but
allowed only the period $P$ and global zero-point $T_{c,0}$ to float freely. 
In addition, to help break the degeneracy between $a/R_\star$ and $b$, we imposed a Gaussian prior
on the stellar density as $\rho_\star=30.5\pm2.7$ g cm$^{-1}$ based on Table \ref{hyo1} in the MCMC analysis. 
We jointly modeled and fitted all the light curve segments from the K2 and TESS data, 
and derived the global posterior distribution for the fitting parameters.

%%%%%%%%%%%%%%%%%%%
\begin{figure}
\centering
\includegraphics[width=8.5cm]{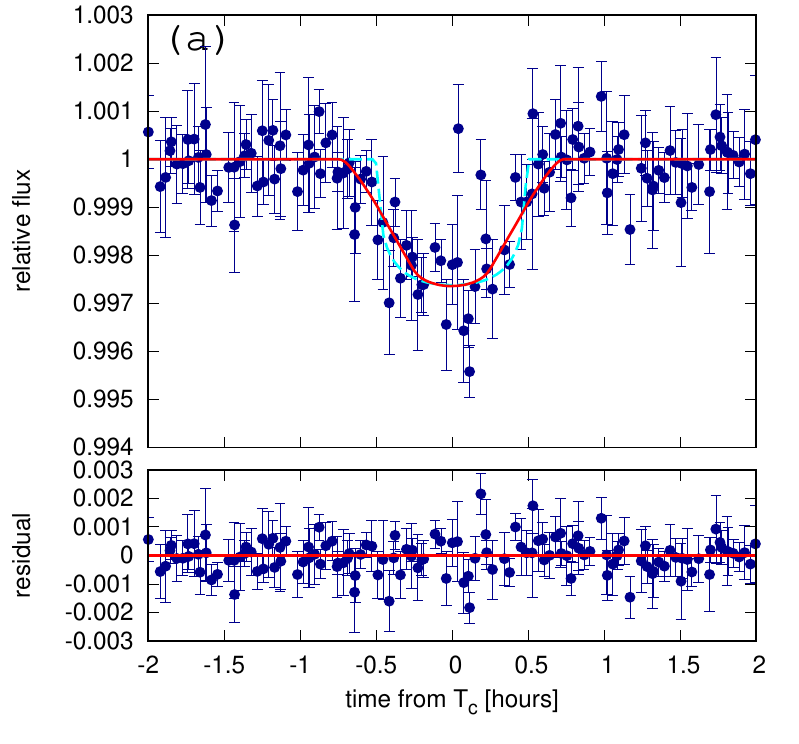}
\includegraphics[width=8.5cm]{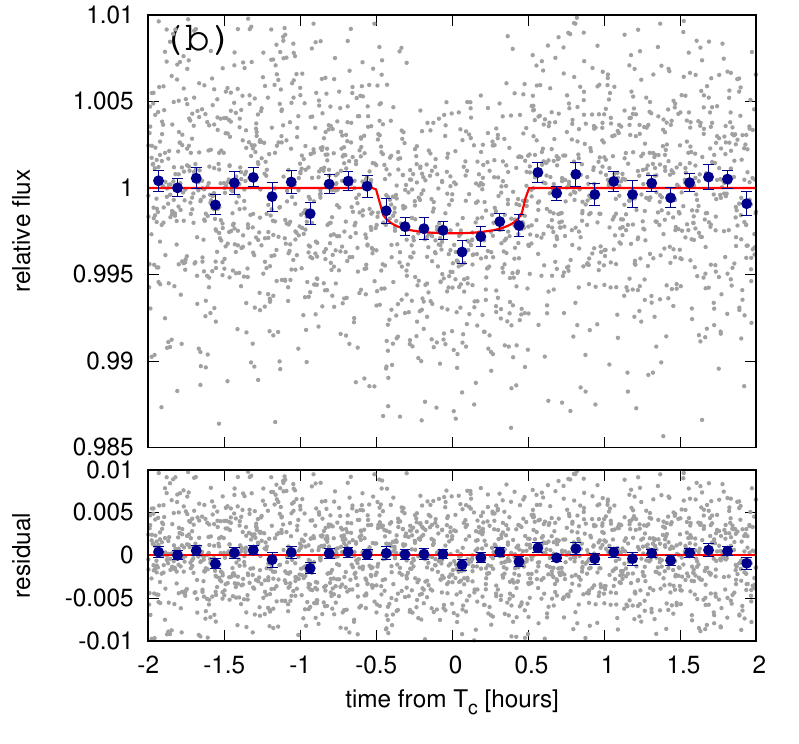}
\caption{
Folded K2 (panel (a)) and TESS (panel (b)) transit light curves of \target b, for 
which a linear ephemeris is assumed. In both panels, red solid lines indicate the
best-fit transit model (the $e$-prior fit) for the observed data, 
and flux residuals from the model are plotted
at the bottom. For panel (a), K2's long cadence is taken into account in drawing the best-fit theoretical curve, and the original transit model before binning is shown by the cyan dashed line in the same panel. 
}
\label{fig:folded}
\end{figure}
%%%%%%%%%%%%%%%%%%%
In this joint analysis, we tried two different fits: one with $e=0$ and the other with a floating $e$. 
The result of the $e=0$ model fit is presented in the third column of Table \ref{hyo2}. 
The derived parameters are in good agreement with the ones derived from fitting the K2
or TESS light curve alone, but with much smaller uncertainties. 
For the floating $e$ model, we first attempted an MCMC analysis allowing $\sqrt{e}\cos\omega$ and 
$\sqrt{e}\sin\omega$ to vary freely, but we found the fit did not converge likely owing to the
degeneracy in the fitting parameters together with the shallow transit and very short transit duration \citep[see e.g.,][]{2022A&A...667A...1S}. 
%(only $\approx 1$ hour). 
Orbital eccentricities of 
close-in planets have been studied in the past, and low or moderate eccentricities were suggested for small, close-in planets \citep[e.g.,][]{2011arXiv1109.2497M, 2019AJ....157...61V}. 
Thus, we imposed weak Gaussian priors on $\sqrt{e}\cos\omega$ and $\sqrt{e}\sin\omega$ 
with the mean and standard deviation of 0 and 0.335, respectively, so that the
$1\sigma$ upper limit of $e$ becomes 0.45 \citep{2011arXiv1109.2497M}. 
These priors were used only to account for the uncertainty in $e$ and derive realistic
errors for the other fitting parameters. 
The result of this MCMC fit for the $e\neq 0$ model is shown in the rightmost column of Table \ref{hyo2}. 
The phase-folded transit light curves after corrections for the out-of-transit baseline 
are plotted in Figure \ref{fig:folded} along with the best-fit transit models.

\subsection{Statistical Validation of \target b} \label{sec:validation}

We computed the false positive probability (FPP) of \target b using the Python package {\tt vespa} \citep{Morton2015}, which was developed for the statistical validation of planets from the {\it Kepler} mission \citep{Morton2016}. {\tt vespa} employs a robust statistical framework to compare the likelihood of the planetary scenario to the likelihoods of several astrophysical false positive scenarios involving eclipsing binaries, relying on simulated eclipsing populations based on the {\tt TRILEGAL} Galaxy model \citep{Girardi2005}. As inputs to {\tt vespa}, we used an exclusion radius of 4\arcsec, WIYN/NESSI contrast curves (Section~\ref{sec:speckle}), phase-folded TESS photometry and a $3\sigma$ upper limit on the secondary eclipse depth (Section~\ref{sec:ktwo}), as well as broadband optical and NIR photometry, the Gaia parallax, and our spectroscopically-derived estimates of the effective temperature and metallicity of the host star (Table~\ref{hyo1}). The FPP from {\tt vespa} for \target b is 2$\times$10$^{-4}$, well below the commonly used validation threshold of 1\%, which is in good agreement with the non-detection of any massive companions in our high resolution imaging and RV measurements. 

To confirm the low FPP, we also implemented another Python package, {\tt TRICERATOPS} \citep{2021AJ....161...24G}, which was developed to vet and validate TESS planet candidates. {\tt TRICERATOPS} also returned a low FPP of $0.15\,\%$ for \target b, 
confirming the result by {\tt vespa}. 
\target b is therefore incompatible with any known false positive scenarios, so we conclude that it has to be a real planet.

\subsection{Analyses of IRD Data} \label{sec:RV}

\begin{figure}
\centering
\includegraphics[width=8.5cm]{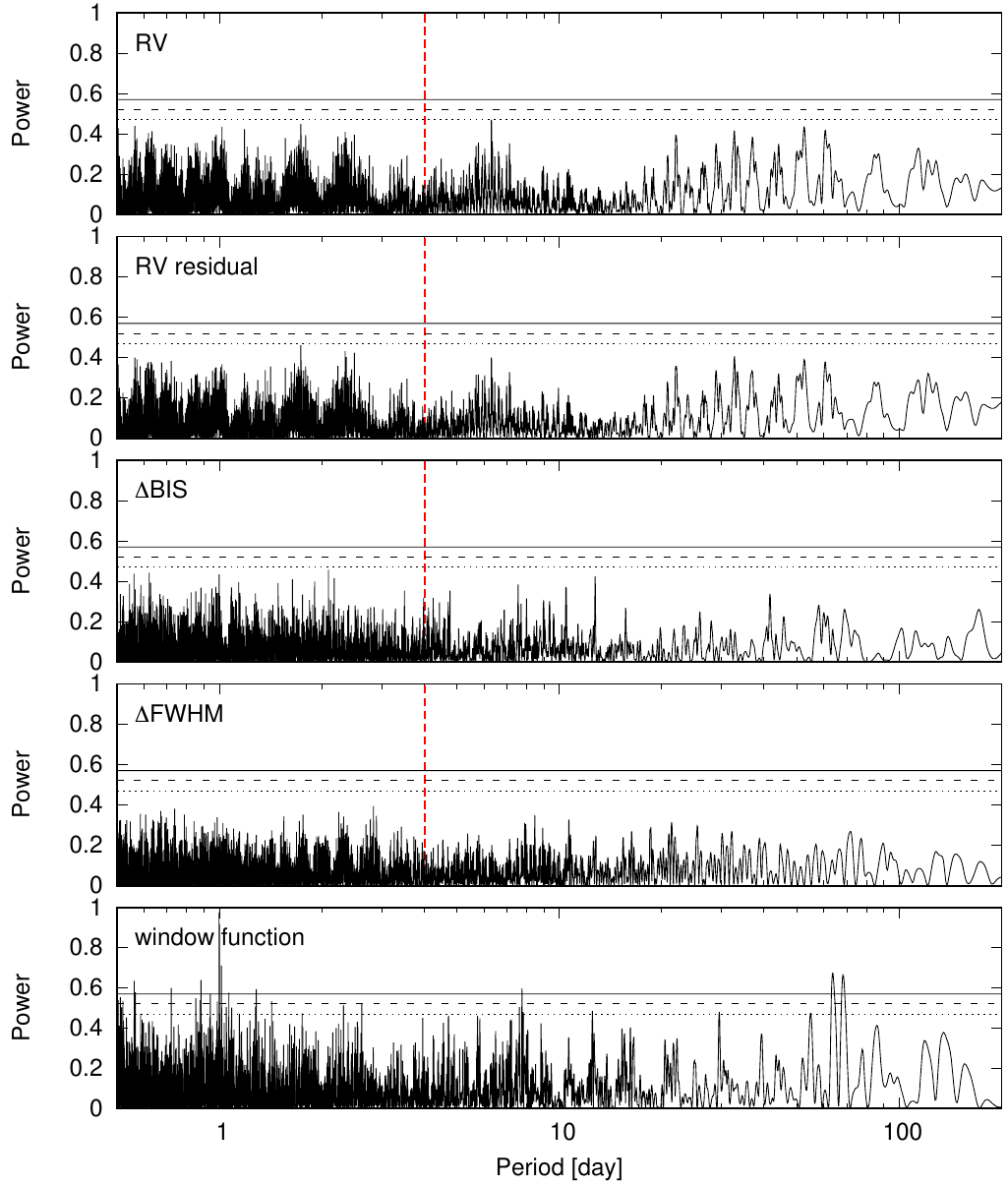}
\caption{LS periodograms for the observed RVs and spectral indices 
($\mathrm{\Delta FWHM}$ and $\mathrm{\Delta BIS}$) extracted from the IRD spectra. 
In each panel, the vertical dashed line represents the orbital period of \target b. 
The three horizontal lines correspond to the FAPs of 5\,\%, 1\,\%, and 0.1\,\%
from the bottom, respectively. The second panel from the top shows the LS power
for the RV residuals after subtracting the best-fit orbit of \target b. 
}
\label{fig:ird-periodogram}
\end{figure}

%In this subsection, we describe the analyses of RV data obtained by Subaru/IRD. 
RV data presented in Table \ref{tab:ird} shows a relatively large scatter with 
a root-mean-square 
(RMS) of 11.9 m s$^{-1}$, which is significantly larger than the mean RV
internal error of 6.5 m s$^{-1}$. The expected RV variation by \target b is only 
a few m s$^{-1}$ based on the mass-radius relation \citep[e.g.,][]{{2020A&A...634A..43O}}. 
In order to see if we can detect any planet signal associated with the period of
\target b, we computed the generalized LS periodogram \citep{2009A&A...496..577Z} 
for the observed RV data. 
However, no significant peak showed up at $P=4.02$ days, as drawn in Figure \ref{fig:ird-periodogram} (more generally, no peak exceeded the $\mathrm{FAP}=1\,\%$ threshold). 
While a fraction of the excess RV
scatter might be ascribed to the instrumental systematics of IRD \citep[e.g.,][]{2021AJ....162..161H, 2022PASJ...74..904H}, one should
notice that \target\ has a relatively short period of rotation ($P_\mathrm{rot}=4.1-4.5$ days, 
according to the LS periodograms) and both K2 and TESS light curves exhibit 
significant variations due to star spots. 
The stellar rotation period and the radius translate to the equatorial rotation 
velocity of $\approx 2.3$ km s$^{-1}$, which places an upper limit on $v\sin i$ of
\target. Assuming that the system has a spin-orbit alignment (i.e., $v\sin i\approx 2.3$ km s$^{-1}$)
and that the star has an effective surface-spot area of $\approx 0.2-0.4\,\%$ (from the light curves
in Figure \ref{fig:normalizedLC}), one can roughly estimate the expected RV jitter amplitude as $8-16$ m s$^{-1}$ \citep[e.g.,][]{2007A&A...473..983D, 2012A&A...545A.109B}, which is comparable to the amplitude of the excess RV scatter. 
Therefore, in the following discussion, we model the observed RVs taking into account
the spot-induced RV jitters, to obtain an accurate constraint on the mass of \target b. 

%%%%%%%%%%%%%%%%%%%
%%%%%%%%%%%%%%%%%%%

To model the observed RVs, we employed the Gaussian-process (GP) based approach 
described in \citet{2015MNRAS.452.2269R}; in short, they applied the GP regression 
to the observed RVs
together with the auxiliary parameters (the $\log R_\mathrm{HK}^\prime$ index and 
the inverse slope of the cross-correlation bisector: BIS) measured from the same spectra. 
To implement a similar GP regression, we measured some activity indices from the
IRD spectra in a similar manner to those in \citet{2022PASJ...74..904H}. 
Since the Ca HK line is not covered by IRD, instead of $\log R_\mathrm{HK}^\prime$
we adopted the FWHM of the mean line profile, which is known to display a similar behavior 
to $\log R_\mathrm{HK}^\prime$ \citep{2015MNRAS.452.2269R}. 
The mean line profile for each IRD spectrum is extracted using the least-squares
deconvolution method \citep[LSD:][]{1997MNRAS.291..658D, 2015A&A...583A..51A}, and in doing so 
we used only OH lines, which are the most 
dominant opacity sources in the $H-$band spectra of early-mid M 
dwarfs\footnote{We did not use $YJ-$band spectra for activity-index measurements,
since a significant fraction of IRD data were taken with the $H-$band detector alone
(see Section \ref{sec:obs-ird}).}. 
From the same LSD profile, we also measured BIS. 
For both FWHM and BIS measurements, 
we subtracted the temporal mean values for individual lines and averaged over
many different lines in the same spectrum to focus on the relative variations in FWHM and BIS,
%in km s$^{-1}$, 
which we call $\Delta\mathrm{FWHM}(t)$ and $\Delta\mathrm{BIS}(t)$. 
More details of activity-index measurements for IRD spectra will be presented in our
upcoming papers. 
%(e.g., Hirano et al. in prep.). 
The time series and LS periodograms of $\Delta\mathrm{FWHM}(t)$ and $\Delta\mathrm{BIS}(t)$ 
are also presented in Table \ref{tab:ird} and Figure \ref{fig:ird-periodogram}, respectively.

Following \citet{2015MNRAS.452.2269R}, we model the stochastic component of the observed RVs ($\Delta\mathrm{RV}(t)$) as well as $\Delta\mathrm{FWHM}(t)$ and 
$\Delta\mathrm{BIS}(t)$ by the following equations:
%%%%%%%%%%%%%%%%%%%
\begin{eqnarray}
\label{eq:GPmodel}
\Delta\mathrm{RV}(t)&=& V_cG(t) + V_r\dot{G}(t),\\
\Delta\mathrm{FWHM}(t)&=&WG(t),\\
\Delta\mathrm{BIS}(t)&=& B_cG(t)+ B_r\dot{G}(t), 
\end{eqnarray}
%%%%%%%%%%%%%%%%%%%
where $G(t)$ is a latent function associated with the fractional spot area and its derivative. The coefficients $V_c$, $V_r$, $W$, $B_c$, and $B_r$ are determined by fitting the observed data. In a multi-dimensional GP regression, fitting parameters are optimized so that the logarithm of the likelihood function $\mathcal{L}$ is maximized:
%%%%%%%%%%%%%%%%%%%
\begin{eqnarray}
\label{eq:likelihood}
\log \mathcal{L} = -\frac{1}{2}(\mathbf{y}-\mathbf{m})^\mathrm{T}\Sigma^{-1}(\mathbf{y}-\mathbf{m})\nonumber\\
-\frac{1}{2}\log(|\Sigma|)-\frac{n}{2}\log(2\pi),
\end{eqnarray}
%%%%%%%%%%%%%%%%%%%
where $\mathbf{y}$ and $\mathbf{m}$ are the vectors (with $n$ components) representing 
the observed variables and mean functions, respectively, and $\Sigma$ is the covariant 
matrix of the input variables. In our formulation, $n=3N_\mathrm{data}$ with $N_\mathrm{data}$
being the number of IRD spectra ($=42$ in the present case). 
We adopt the mean functions $\mathbf{m}$ given by 
%%%%%%%%%%%%%%%%%%%
\begin{eqnarray}
\label{eq:mean}
m_\mathrm{RV}(t)&=&K\{\cos(f+\omega)+e\cos\omega\}+\gamma_\mathrm{RV},\\
m_{\Delta\mathrm{FWHM}}(t)&=&\gamma_\mathrm{\Delta FWHM},\\
m_{\Delta\mathrm{BIS}}(t)&=&\gamma_\mathrm{\Delta BIS}, 
\end{eqnarray}
%%%%%%%%%%%%%%%%%%%
where $f$ is the true anomaly and $K$ is the RV semi-amplitude due to \target b's Keplerian motion.  The constant offset $\gamma$ in each variable is also optimized in the fit. 
For the RV modeling, a fixed orbital period $P$ is used based on the transit light curve 
analyses in Section \ref{sec:ana_photometry}.

The only remaining quantity to be determined {\it a priori} is the form of the covariant matrix $\Sigma$. 
%Given that the light curves exhibit flux modulations likely due to spots, which seem to evolve on a timescale of $\sim 10$ days, 
Here, we adopt the quasi-periodic kernel for the covariance between the latent functions $G(t_i)$ 
and $G(t_j)$:
 %%%%%%%%%%%%%%%%%%%
\begin{eqnarray}
\label{eq:quasi-per2}
k_\mathrm{qp}(t_i,t_j) \equiv \exp\left[ -\frac{\sin^2\{\pi(t_i-t_j)/P_\mathrm{rot}^\prime\}}{2\lambda_\mathrm{p}^2} -\frac{(t_i-t_j)^2}{2\lambda_\mathrm{e}^2} \right], ~~
\end{eqnarray}
%%%%%%%%%%%%%%%%%%%
and compute the covariant matrix $\Sigma$ following the expressions given by \citet{2015MNRAS.452.2269R}. 
The hyperparameters $P_\mathrm{rot}^\prime$, $\lambda_\mathrm{p}$, and $\lambda_\mathrm{e}$
are also fitting parameters in the regression.

%%%%%%%%%%%%%%%%%%%%%%%%%%%%%%%%%%%%%%%%%%%%%%%%%%%%%%%%%%%%%%%%%%%%%%
\begin{table}[t]
%\tabletypesize{\small}
\centering
\caption{GP regressions to the Light Curves}\label{hyo3}
\begin{tabular}{lcc}
\hline\hline
Data Set & K2 & TESS \\\hline
(Fitting Parameter) & &  \\
$A$ & $0.00288_{-0.00031}^{+0.00039}$ & $0.00132_{-0.00013}^{+0.00017}$\\
$P_\mathrm{rot}^\prime$ (days)& $4.164_{-0.065}^{+0.084}$ & $4.54_{-0.11}^{+0.10}$ \\
$\lambda_\mathrm{p}$ & $0.692_{-0.058}^{+0.069}$ & $0.411_{-0.055}^{+0.069}$ \\
$\lambda_\mathrm{e}$ (days)& $4.71\pm 0.23$ & $3.65_{-0.39}^{+0.41}$ \\
\hline
\end{tabular}
\end{table}
%%%%%%%%%%%%%%%%%%%%%%%%%%%%%%%%%%%%%%%%%%%%%%%%%%%%%%%%%%%%%%%%%%%%%%

We attempted to implement a multi-dimensional GP regression to the observed quantities (RV, $\Delta\mathrm{FWHM}(t)$, 
and $\Delta\mathrm{BIS}(t)$) using our custom MCMC code \citep{2016ApJ...820...41H}, in which all the fitting parameters 
are allowed to float freely, but we immediately
found that the fit did not converge, often resulting in walkers being stuck to different local minima. 
This is likely attributable to the small number of data points (only 42), spread over many years; In particular, the sparseness of the data seemed to prohibit reliable estimations of the periodicity ($P_\mathrm{rot}^\prime$) and evolution timescale ($\lambda_\mathrm{e}$) of RV jitters. 
We thus resorted to a two-step approach employed by \citet{2015ApJ...808..127G}, who used 
the light curves to constrain the hyperparameters in the covariance kernel (Equation (\ref{eq:quasi-per2})), 
and performed GP regressions to the K2 and TESS data before analyzing the spectral data. 
Adopting the covariance matrix for the flux values, whose elements are given by 
%%%%%%%%%%%%%%%%%%%
\begin{eqnarray}
\label{eq:quasi-per}
\Sigma_{ij} = A^2 k_\mathrm{qp}(t_i,t_j) + \sigma_i^2 \delta(t_i-t_j), 
\end{eqnarray}
%%%%%%%%%%%%%%%%%%%
where $\sigma_i$ is the $i-$th flux error, 
we ran MCMC analyses to estimate $P_\mathrm{rot}^\prime$, $\lambda_\mathrm{p}$, 
and $\lambda_\mathrm{e}$ as well as the correlation amplitude $A$ for each of K2 and TESS
light curves (binned after masking the transits). 
The results of these analyses are presented in Table \ref{hyo3}, and GP regressions to 
the light curves are plotted by the red solid lines in Figure \ref{fig:normalizedLC}. 
The derived parameters show a moderate disagreement ($\approx 3\,\sigma$ except $A$) between K2 and TESS data, but these are likely due to different properties of the two data sets (e.g., different observing bands) as well as data processing. The GP-based period ($P_\mathrm{rot}^\prime\approx4.2-4.5$ days) roughly agrees with the rotation period derived from the LS periodograms ($P_\mathrm{rot}\approx 4.3-4.4$ days; see Section \ref{sec:parameters}), which justifies the use of the quasi-periodic kernel for the GP regression.

%%%%%%%%%%%%%%%%%%%%%%%%%%%%%%%%%%%%%%%%%%%%%%%%%%%%%%%%%%%%%%%%%%%%%%
\begin{table}[t]
%\tabletypesize{\small}
\centering
\caption{GP regressions to the IRD Data (RV + Spectral Indices). Guassian priors are imposed on the fitting parameters with asterisk. }\label{hyo4}
\begin{tabular}{lcc}
\hline\hline
Condition & $e=0$ & $e$ prior \\\hline
(Fitting Parameter) & &  \\
$K$ (m s$^{-1}$) & $ 4.4_{-3.3}^{+3.1} $ & $ 4.1_{-3.6}^{+3.5} $ \\
$\sqrt{e}\cos\omega$$^*$ & 0 (fixed) & $ 0.03 \pm 0.29 $ \\
$\sqrt{e}\sin\omega$$^*$ & 0 (fixed) & $ 0.00 \pm 0.32 $ \\
$\gamma_\mathrm{RV}$ (m s$^{-1}$) & $ 1.6\pm2.4 $ & $ 1.6_{-2.5}^{+2.4} $ \\
$\gamma_{\Delta \mathrm{FWHM}}$ (km s$^{-1}$) & $ -0.038\pm0.018 $ & $ -0.038_\pm 0.018 $ \\
$\gamma_{\Delta \mathrm{BIS}}$ (km s$^{-1}$) & $ 0.098_{-0.042}^{+0.043} $ & $ 0.096\pm 0.042 $ \\
$P_\mathrm{rot}^\prime$$^*$ (days)& $ 4.295_{-0.060}^{+0.061} $ & $ 4.298\pm 0.061 $ \\
$\lambda_\mathrm{p}$$^*$ & $ 0.564\pm 0.044 $ & $ 0.562\pm 0.044 $ \\
$\lambda_\mathrm{e}$$^*$ (days)& $ 4.46\pm 0.20$ & $ 4.46 \pm 0.20 $ \\
$V_c$ (m s$^{-1}$) & $ 5.0_{-3.3}^{+3.4} $ & $ 5.2 \pm 3.4 $ \\
$V_r$ (m s$^{-1}$) & $ -2.3_{-3.5}^{+4.2} $ & $ -2.5_{-3.5}^{+4.5} $ \\
$W$ (km s$^{-1}$) & $ 0.008_{-0.049}^{+0.041} $ & $ 0.008_{-0.048}^{+0.039} $ \\
$B_c$ (km s$^{-1}$) & $ 0.00_{-0.13}^{+0.12} $ & $ -0.01_{-0.13}^{+0.12} $ \\
$B_r$ (km s$^{-1}$) & $ 0.104_{-0.052}^{+0.042} $ & $ 0.102_{-0.053}^{+0.043} $ \\
$C_\mathrm{RV}$ & $ 1.56_{-0.24}^{+0.31} $ & $ 1.55_{-0.24}^{+0.30} $ \\
$C_\mathrm{\Delta FWHM}$ & $ 1.93_{-0.25}^{+0.27} $ & $ 1.94_{-0.25}^{+0.27} $ \\
$C_\mathrm{\Delta BIS}$ & $ 0.61_{-0.10}^{+0.15} $ & $ 0.61_{-0.10}^{+0.15} $ \\
\hline
\end{tabular}
\end{table}
%%%%%%%%%%%%%%%%%%%%%%%%%%%%%%%%%%%%%%%%%%%%%%%%%%%%%%%%%%%%%%%%%%%%%%

%%%%%%%%%%%%%%%%%%%
\begin{figure*}
\centering
\includegraphics[width=15cm]{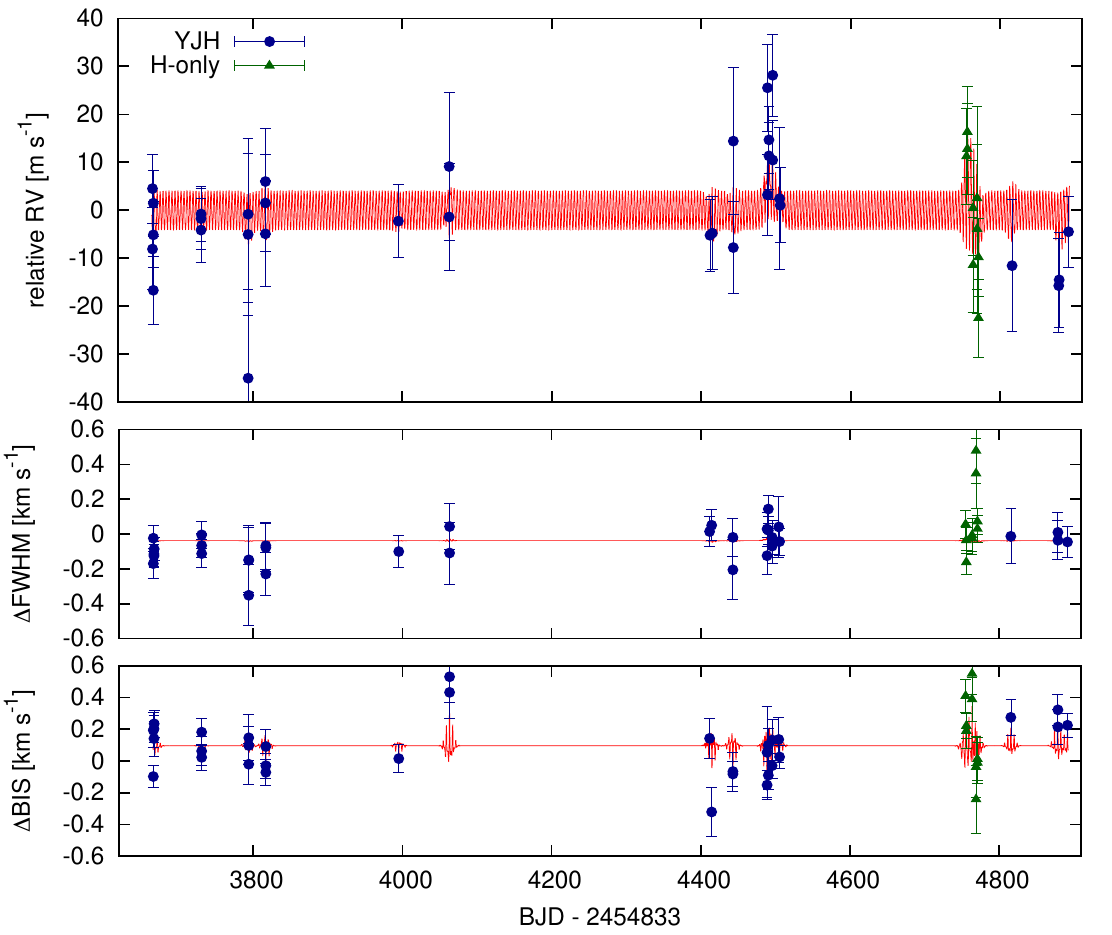}
\caption{Observed RVs (top) as well as the activity indicators (middle and bottom panels)
measured from the IRD spectra. The red solid lines indicate the multi-dimensional GP regressions to the observed quantities using the quasi-periodic GP kernel. 
}
\label{fig:IRDtime}
\end{figure*}
%%%%%%%%%%%%%%%%%%%

We computed the weighted mean for each hyperparameter in Table \ref{hyo3}, and used this result from the light curve analysis 
as priors in GP regressions for the spectral data; employing Gaussian priors of $P_\mathrm{rot}^\prime=4.291\pm 0.061$ days, 
$\lambda_\mathrm{p}=0.554\pm 0.044$, and $\lambda_\mathrm{e}=4.45\pm 0.20$ days,
we implemented the MCMC fit to the observed RVs along with $\Delta\mathrm{FWHM}$ and $\Delta\mathrm{BIS}$. 
In the analysis, we introduced additional fitting parameters 
$C_\mathrm{RV}$, $C_\mathrm{\Delta FWHM}$, and $C_\mathrm{\Delta BIS}$
to take into account underestimation/overestimation for
internal errors of input variables, which are inserted into the diagonal components of the covariance matrix $\Sigma$ as
%%%%%%%%%%%%%%%%%%%
\begin{eqnarray}
C_l^2 \sigma_i^{(l)2}\delta(t_i-t_j),
\end{eqnarray}
%%%%%%%%%%%%%%%%%%%
where $l=\{\mathrm{RV}, \mathrm{\Delta FWHM}, \mathrm{\Delta BIS}\}$ and $\sigma_i^{(l)}$ is the 
input statistical error at time $t_i$. 
We employed this formulation to model the white noise components in the fit, as opposed to adding extra terms of white noise in quadrature, as we found the input statistical errors of BIS we measured from the spectral analysis are slightly overestimated (i.e., $C_{\Delta\mathrm{BIS}}<1.0$).

As in the case of transit photometry analyses, we tried two different fits with $e=0$ and 
with priors on $e$; A fit with completely free $\sqrt{e}\cos\omega$ and $\sqrt{e}\sin\omega$ did not result in a
meaningful constraint on $e$ likely due to the small number of data points and significant RV jitters, 
and hence we imposed Gaussian priors on these parameters as in Section \ref{sec:ana_photometry}
only to obtain realistic uncertainties for the other fitting parameters. 
All the fitting parameters other than $P_\mathrm{rot}^\prime$, $\lambda_\mathrm{p}$, 
$\lambda_\mathrm{e}$, $\sqrt{e}\cos\omega$, and $\sqrt{e}\sin\omega$ being allowed to 
vary freely, we performed GP regressions to the spectral data by implementing MCMC analyses
\citep{2016ApJ...820...41H}. 
The results of these analyses are summarized in Table \ref{hyo4}.

%%%%%%%%%%%%%%%%%%%
\begin{figure}
\centering
\includegraphics[width=8.5cm]{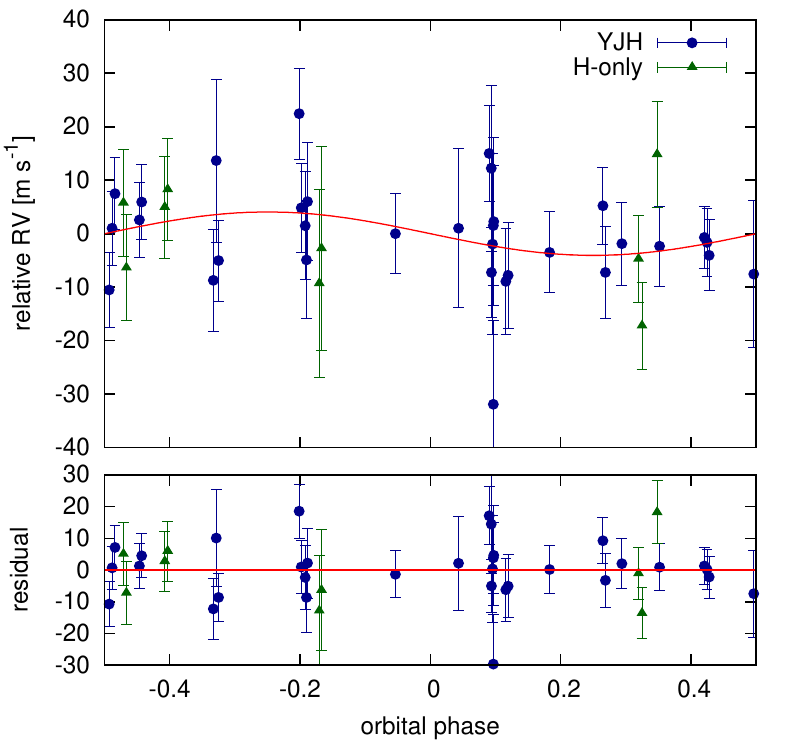}
\caption{Phase-folded RV curve after subtracting the correlated noise component by the GP regression. The red solid line represents the best-fit Keplerian orbit of \target b, and
the bottom panel plots the residuals from the best-fit orbit. 
}
\label{fig:IRDphase}
\end{figure}
%%%%%%%%%%%%%%%%%%%

Figure \ref{fig:IRDtime} plots the time series of the observed RVs along with the two spectral indices, 
and the red solid line for each panel indicates the best-fit GP regression (with the $e$ prior) to the observed data. 
The phase-folded RV curve after removing the GP stochastic components is given in Figure \ref{fig:IRDphase}. 
The derived RV semi-amplitude $K$ of $4.1_{-3.6}^{+3.5}$ m s$^{-1}$ well agrees with that expected from 
the mass-radius relation \citep[$K=1-2$ m s$^{-1}$;][]{2020A&A...634A..43O}, but it is also compatible with a non-detection of the planet within 
$1.1\,\sigma$, which is consistent with the lack of a significant peak at the orbital period in the LS periodogram of the RV data. 
Nonetheless, the $1\,\sigma$ upper limit of $K$ translates to the planet mass constraint of $<5.7\,M_\oplus$, 
which, along with the lack of a long-term RV trend, allows us to completely rule out FP with a stellar companion.

%%%%%%%%%%%%%%%%%%%%%%%%%%

\section{Discussion} \label{sec:discussion}

%%%%%%%%%%%%%%%%%%%%%%%%%%%%%%%%%%%%%%%%%%%%%%%%%%%%%%%%%%%%%%%%%%%%%%
\begin{table}[t]
%\tabletypesize{\small}
%\begin{center}
\centering
\caption{Final Planetary Parameters of \target b}\label{hyo5}
\begin{tabular}{lc}
\hline\hline
Parameter & Value \\\hline
$P$ (days) & $4.0179694\pm 0.0000027$ \\
$T_{c,0}-2454833$ (BJD) & $3267.07116\pm 0.00069$ \\
$R_p$ ($R_\oplus$) & $1.015 \pm 0.051$\\
$M_p$ ($M_\oplus$) & $3.0 \pm 2.7$\\
%$\rho_p$ (g cm$^{-1}$) & \\
$a$ (au) & $0.0270 \pm 0.00023$\\
$i_o$ (deg) & $89.32 \pm 0.41$\\
$S_p$ ($S_\oplus$) & $4.82_{-0.42}^{+0.45}$\\
$T_\mathrm{eq}$ ($A_B=0$) (K) & $412.4 \pm 9.3$\\
$T_\mathrm{eq}$ ($A_B=0.3$) (K) & $377.2 \pm 8.6$\\
\hline
\end{tabular}
%\end{center}
\end{table}
%%%%%%%%%%%%%%%%%%%%%%%%%%%%%%%%%%%%%%%%%%%%%%%%%%%%%%%%%%%%%%%%%%%%%%

Adopting the parameters derived in Section \ref{sec:ana} (for the $e$ prior fit), we calculated the final planetary parameters of \target b, including the physical planet radius $R_p$, 
mass $M_p$, semi-major axis $a$, orbital inclination $i_o$, stellar insolation onto the
planet $S_p$, and equilibrium 
temperature $T_\mathrm{eq}$, for which we adopted two Bond albedos:
$A_B=0$ and $A_B=0.3$. The derived planet parameters are summarized in Table \ref{hyo5};
\target b's mass is estimated to be $3.0\pm2.7\,M_\oplus$ ($M_p<7.5\,M_\oplus$ at $95\,\%$
confidence). 

%%%%%%%%%%%%%%%%%%%
\begin{figure}
\centering
\includegraphics[width=8.5cm]{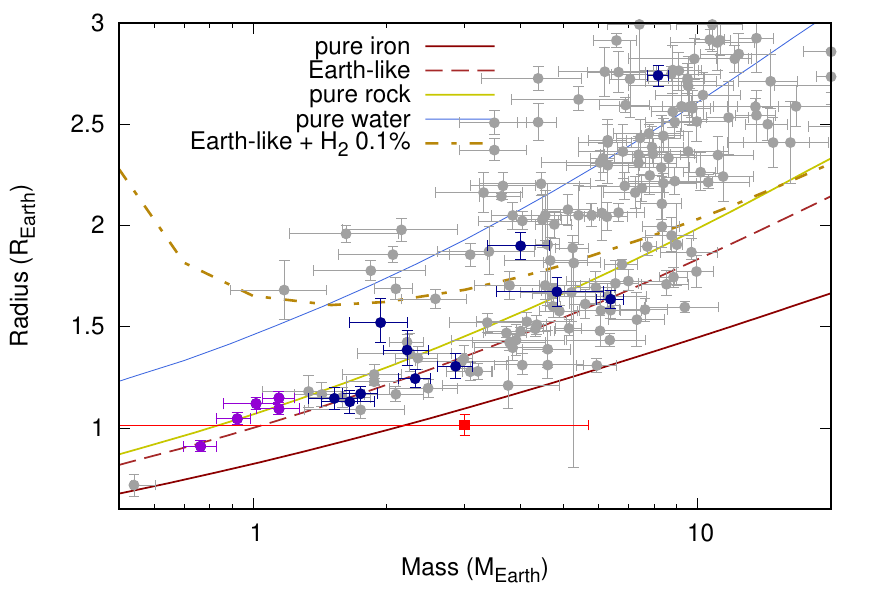}
\caption{
Mass-radius diagram for well-characterized transiting planets, taken from the TEPcat catalog \citep{2011MNRAS.417.2166S}. 
\target b, planets around the low-mass stars ($0.1\,M_\odot<M_\star<0.3\,M_\odot$), and 
TRAPPIST-1 planets are plotted by the red square, dark-blue circles, and purple circles, respectively. 
Theoretical curves are drawn using the models by \citet{2016ApJ...819..127Z, 2019PNAS..116.9723Z}, 
in which an equilibrium temperature of 400 K was assumed. 
}
\label{fig:MR}
\end{figure}
%%%%%%%%%%%%%%%%%%%

To compare \target b with other small planets around the lowest mass stars, 
we were tempted to put \target b in the mass-radius diagram, although the 
mass constraint for the planet is rather weak. 
Figure \ref{fig:MR} plots the planet masses versus radii for the well-characterized 
planets based on the TEPcat database \citep{2011MNRAS.417.2166S};
\target b and the planets around the lowest mass stars ($<0.3\,M_\odot$)
are plotted by the red square and colored circles, respectively. 
As expected, due to the weak constraint on the planet mass, we are unable to 
discuss in detail the internal structure of \target b at this point. 
Nonetheless, the planet seems to be in line with the trend of other known small planets 
around the lowest-mass stars, having an Earth-like composition. 
The expected mass for a $100\,\%$-iron planet with $R_p=1.015\,R_\oplus$ is $\approx 2.2\,M_\oplus$, which gives a theoretical upper limit on the mass of 
\target b (i.e., its true mass is likely less than this limit). 
It is also evident from the figure that the size of the planet is incompatible 
with possessing a H$_2$-dominated atmosphere with $T_\mathrm{eq}=400$ K. 
According to the recent finding by \citet{2022Sci...377.1211L} on the population of small planets around low-mass stars, a $1\,R_\oplus$ planet has to have a rocky composition as opposed to water-rich planets having slightly larger radii ($\gtrsim 1.5\,R_\oplus$). 

In theoretical aspects, 
%\textcolor{red}{
recent models of planet formation via planetesimal accretion \citep[][]{Kimura+2022} predict that there exist rocky planets of $\sim 0.3-2\,M_\oplus$ with no atmosphere and no or small amounts of water around late M dwarfs. Alternatively, in the context of the pebble accretion scenario, given our estimated planetary mass is comparable to
%} 
%Note that in the context of the planet formation scenario, the estimated mass is consistent with 
the pebble isolation mass ($\sim 1-2\,M_\oplus$) %for planets 
near the snowline around low-mass stars \citep{2019A&A...632A...7L, 2020A&A...638A..88L}, %suggesting that the pebble accretion and subsequent 
%\textcolor{red}{
inward migration of planetary embryos with the isolation mass followed by
%} 
giant impacts \citep[e.g.,][]{2020A&A...642A..23M} %may have played some roles in its formation. 
%\textcolor{red}{
would be another possible formation pathway for \target b.
%}

At 22 pc, \target\ is currently the closest planet-hosting star from Earth identified by the Kepler/K2 missions, followed by K2-129 \citep[27.8 pc;][]{2017AJ....154..207D} and G 9-40 \citep[27.9 pc;][]{2020AJ....159..100S}. 
\target\ is also one of the lowest mass stars hosting an Earth-like planet ($R_p<1.25\,R_\oplus$). 
The only M dwarfs cooler than \target\ hosting Earth-like transiting planets are 
TRAPPIST-1 \citep[12 pc;][]{2017Natur.542..456G}, 
LP 791-18 \citep[26 pc;][]{2019ApJ...883L..16C}, LHS 1140 \citep[15 pc;][]{2017Natur.544..333D}, and Kepler-42 \citep[40 pc;][]{2012ApJ...747..144M}. 
%, according to NASA Exoplanet Archive\footnote{\url{https://exoplanetarchive.ipac.caltech.edu/index.html}}. 
Moreover, as of today, there are only ten transiting-planet (of any size) hosting stars cooler than \target, seven of which are more distant stars than \target.

Given its proximity to Earth and moderate transit depth ($\approx 0.25\,\%$), 
\target b could be a potential target for future atmospheric characterizations of Earth-like planets, 
especially those having a relatively low temperature (e.g., $T_\mathrm{eq}<500$ K). 
%The prospect for future atmospheric characterizations, such as
%the transmission spectroscopy metric \citep[TSM;][]{2018PASP..130k4401K}, 
%will be discussed in detail once the mass (or scale height) of the planet is pinned down 
%by future RV observations. 
To determine the prospects for atmospheric characterization, it is necessary to constrain the planet mass 
more precisely via further RV observations. Existing spectrographs for precision RV measurements can potentially place a good constraint on the planet mass, but with the current RV precision achieved by IRD ($3-4$ m s$^{-1}$), a huge number of RV measurements ($>200$) would be required. Instead, red-optical Doppler instruments such as MAROON-X \citep{2020SPIE11447E..1FS}, which can achieve a better RV precision for early-mid M dwarfs, may be able to deliver a better constraint on \target b's mass with a more reasonable number of observations; for instance, according to the integration time calculator\footnote{\url{https://www.gemini.edu/instrumentation/maroon-x/exposure-time-estimation}}, MAROON-X would be able to achieve peak S/N ratios of $40-50$ (Blue arm) and $\approx 80$ (Red arm) with a 30-minute integration for \target, which translate to an RV precision of $1.2-1.3$ m s$^{-1}$. Assuming this precision and one Earth mass for \target b, we simulated RV observations and analyses to see to what extent one can constrain the mass of the planet by MAROON-X. Since the star exhibits a moderate RV jitter, the mass constraint would ultimately rely on the degree of substantive suppression of RV jitter by post-processing (such as the GP regression above). Therefore, we simulated RV measurements with three different RV jitter magnitudes ($1.0$ m s$^{-1}$, $2.0$ m s$^{-1}$, $5.0$ m s$^{-1}$), which were included as additional Gaussian scatters to simulated data points\footnote{RV jitters are correlated noise, but we assumed that correlated-noise components are more or less suppressed by post-processing and extending the temporal baseline of RV measurements. Thus, the assumed jitters here are remaining noise after their suppression. }.  
As a result of simulating and analyzing 150 RV points by MAROON-X observations, we found that the planet mass ($M_p\,=\,1.0\,M_\oplus$) is constrained with $6.2\,\sigma$, $4.2\,\sigma$, and $2.0\,\sigma$ for the assumed RV jitter values of $1.0$ m s$^{-1}$, $2.0$ m s$^{-1}$, and $5.0$ m s$^{-1}$, respectively. 

Once the planet mass is measured at the $4-5\,\sigma$ level, one may simulate the atmospheric characterization with space-based telescopes (e.g., JWST) and/or future 20-30m class telescopes. Detailed simulated observations for \target b with existing and future facilities are beyond the scope of this paper, but we briefly mention the feasibility for JWST observations based on the atmospheric scale height of \target b; Assuming an Earth mass, $T_\mathrm{eq}$ of 377 K, and an Earth-like atmosphere (i.e., a mean molecular mass of $4.8\times 10^{-26}$ kg) for \target b, we obtain an expected atmospheric scale height of $H\approx 11$ km. The amplitude of spectral features in transmission spectroscopy \citep[e.g., Equation (3) of][]{2018haex.bookE.100K} is then estimated to be $\approx 16$ ppm, which is comparable to the noise floor ($\sim 10$ ppm) expected for JWST NIRSpec observations \citep{2022ApJ...928L...7R}. Also, given the magnitudes of \target, it would be a challenging target to probe an Earth-like atmosphere, even with JWST. 
However, it does not rule out the possibility of detecting signals for an atmosphere with a higher scale height. 
More detailed prospects for future atmospheric characterization, such as the transmission spectroscopy metric \citep[TSM;][]{2018PASP..130k4401K}, will be discussed once the planet mass is better constrained by future RV monitoring.

\target b is located slightly inner of the classical habitable zone \citep{2016ApJ...819...84K}
%does not reside in the classical habitable zone 
based on the insolation flux onto the planet, 
but an outer planet, if any, in the system with a slightly longer period (e.g., $10-15$ days)
could sit inside the habitable zone. 
%Little is known on the properties of multi-planet systems around lowest-mass stars ($<0.3\,M_\odot$), but assuming that their properties are similar to those in the ``Kepler-multi" systems and the planets have a typical spacing of $\sim 20$ mutual Hill radii \citep{2018AJ....155...48W}, it is quite possible that a secondary planet having $\sim$ Earth mass has an orbital period of $10-15$ days. 
Little is known on the properties of multi-planet systems around the lowest mass stars ($<0.3\,M_\odot$), but assuming that their properties are similar to those in the ``Kepler-multi" systems, the planets could have a typical spacing of $\sim 20$ mutual Hill radii \citep{2018AJ....155...48W}.
Recently, \citet{2023MNRAS.519.2838H} also showed that the typical orbital spacing of planets formed by giant impacts is $\sim 20$ mutual Hill radii independently of stellar masses through N-body simulations.
Thus, it is quite possible that a secondary planet having $\sim 1\,M_\oplus$ has an orbital period of $10-15$ days.

As noted in Section \ref{sec:photometry}, the joint 
BLS analysis for K2 and TESS light curves resulted in a non-detection of any other transiting
objects in the system. 
In order to see if any additional planet signals could 
be detected in the RV data, we ran the periodogram analysis for the IRD RVs 
after eliminating the best-fit orbit of \target b. 
The second panel from the top in Figure \ref{fig:ird-periodogram} draws the LS periodogram 
for the RV residuals; no significant peak is identified. 
The lack of TTVs and significant peaks in the BLS search (Section \ref{sec:photometry}) as well
as the LS periodogram suggests that \target b does not have a ``massive" outer friend, but it does not rule out
the presence of additional low-mass planets in the system; 
An Earth-mass planet ($M_p\sin i_o=1.0\,M_\oplus$) with a period of 10 days would induce 
an RV semi-amplitude of $K=1.0$ m s$^{-1}$, which is well below the detection limit for
the IRD data, especially given the system has relatively large RV jitters. 
Although it is not straightforward to detect an ``Earth-mass" planet with a period of 10 days, 
future RV monitoring would enable us to constrain the presence of an outer planet of a few
Earth masses, in addition to obtaining a better constraint on \target b's mass.

We note that \target b could also be a potential target for the stellar obliquity measurement
by the Rossiter-McLaughlin (RM) effect \citep{1924ApJ....60...15R, 1924ApJ....60...22M}. 
Provided that \target's $v\sin i$ is similar to its equatorial velocity of $\approx 2.3$ km s$^{-1}$\footnote{We attempted to derive $v\sin i$ of \target\ from the IRD spectrum, but
we were unable to constrain it since M5 stars' spectra are dominated by molecular lines, which in many parts disagree with synthetic spectra \citep[e.g.,][]{2013MSAIS..24..128A}. }, 
we expect an RV anomaly semi-amplitude of $3-5$ m s$^{-1}$. 
It is challenging to achieve this RV precision with IRD while obtaining a good time 
sampling during a transit (transit duration is only 1 hour), 
but again, red-optical Doppler
instruments like MAROON-X %\citep{2020SPIE11447E..1FS} 
may be able to detect the RM signal for \target b. 
These RM measurements would also provide observations to look for atmospheric signatures such as in the hydrogen and metal lines.

\section{Summary} \label{sec:summary}

Based on the analysis of K2 light curves, we detected an Earth-sized ($R_p=1.015 \pm 0.051\,R_\oplus$) 
planet candidate with $P=4.02$ days around \target, which is an M5 dwarf star at 22 pc. 
The same target was later observed by TESS, which also identified the same candidate,
calling the system TOI-5557.  
Our follow-up observation campaign between 2018 and 2022, including IRCS AO imaging, WIYN 
speckle observations, and IRD near-infrared spectroscopy, allowed us to 
rule out FP scenarios for \target b and statistically validate the planet. 
The RV measurements by IRD were not capable of pinning down the mass of \target b, 
due to the small number of data points and relatively large RV jitters ($8-10$ m s$^{-1}$), but
our careful GP modeling of the observed RVs together with some activity indicators
($\Delta \mathrm{FWHM}$ and $\Delta \mathrm{BIS}$) 
%placed an upper limit of $5.7\,M_\oplus$ for the mass of the planet. 
placed a weak constraint of $M_p=3.0\pm 2.7\,M_\oplus$ for the mass of the planet. 
The distance of 22 pc and the moderate depth 
of the transit make \target\ a good target for future observations,
including further RV monitoring (e.g., to search for additional planets) and transit spectroscopy. 
%atmospheric characterization. 

%\acknowledgments
\begin{acknowledgements}

We thank the anonymous referee for the insightful comments to improve the manuscript. 
This work was supported by JSPS KAKENHI Grant Numbers JP19K14783, JP21H00035, 
JP18H05442, JP15H02063, JP22000005, JP18H05438, JP21K20388, and JP18H05439, and JST CREST Grant Number JPMJCR1761.
K.W.F.L. was supported by Deutsche Forschungsgemeinschaft grants RA714/14-1 within the DFG Schwerpunkt SPP 1992, Exploring the Diversity of Extrasolar Planets.
R.L. acknowledges funding from University of La Laguna through the Margarita Salas Fellowship from the Spanish Ministry of Universities ref. UNI/551/2021-May 26, and under the EU Next Generation funds.
J.K. gratefully acknowledges the support of the Swedish National Space Agency (SNSA; DNR 2020-00104).
We are grateful to Noriharu Watanabe, Taiki Kagetani, and Yujie Zou for the support of IRD observations. 
We also thank Kento Masuda for the helpful discussion on rotation periods of mid-M dwarfs. 
This work was supported by the KESPRINT collaboration, an international 
consortium devoted to the characterization and research of exoplanets 
discovered with space-based missions (\url{https://kesprint.science/}).
This research is based in part on data collected at Subaru Telescope,
which is operated by the National Astronomical Observatory of Japan.
The data analysis was carried out, in part, on the Multi-wavelength Data Analysis System operated by the Astronomy Data Center (ADC), National Astronomical Observatory of Japan.
Part of this work was carried out at the Jet Propulsion Laboratory, California Institute of Technology, under contract with NASA.
We are honored and grateful for the opportunity of observing the 
Universe from Maunakea, which has the cultural, historical and natural 
significance in Hawaii.

\end{acknowledgements}

\vspace{5mm}
\facilities{Subaru (IRD, IRCS), WIYN (NESSI)}


\software{
%{\tt IRAF} \citep{1993ASPC...52..173T}
{\tt vespa} \citep{Morton2015}, 
{\tt TRICERATOPS} \citep{2021AJ....161...24G}
}



%\clearpage

%\appendix

%\section{Analysis of the Effective RVs} \label{sec:app1}


%\bibliography{hirano2022_reference}{}

\begin{thebibliography}{}
\expandafter\ifx\csname natexlab\endcsname\relax\def\natexlab#1{#1}\fi
\providecommand{\url}[1]{\href{#1}{#1}}
\providecommand{\dodoi}[1]{doi:~\href{http://doi.org/#1}{\nolinkurl{#1}}}
\providecommand{\doeprint}[1]{\href{http://ascl.net/#1}{\nolinkurl{http://ascl.net/#1}}}
\providecommand{\doarXiv}[1]{\href{https://arxiv.org/abs/#1}{\nolinkurl{https://arxiv.org/abs/#1}}}

\bibitem[{{Allard} {et~al.}(2013){Allard}, {Homeier}, {Freytag},
  {Schaffenberger}, {}, \& {Rajpurohit}}]{2013MSAIS..24..128A}
{Allard}, F., {Homeier}, D., {Freytag}, B., {et~al.} 2013, Memorie della
  Societa Astronomica Italiana Supplementi, 24, 128.
\newblock \doarXiv{1302.6559}

\bibitem[{{Asensio Ramos} \& {Petit}(2015)}]{2015A&A...583A..51A}
{Asensio Ramos}, A., \& {Petit}, P. 2015, \aap, 583, A51,
  \dodoi{10.1051/0004-6361/201526401}

\bibitem[{{Bogner} {et~al.}(2022){Bogner}, {Stelzer}, \&
  {Raetz}}]{2022AN....34310079B}
{Bogner}, M., {Stelzer}, B., \& {Raetz}, S. 2022, Astronomische Nachrichten,
  343, e10079, \dodoi{10.1002/asna.20210079}

\bibitem[{{Boisse} {et~al.}(2012){Boisse}, {Bonfils}, \&
  {Santos}}]{2012A&A...545A.109B}
{Boisse}, I., {Bonfils}, X., \& {Santos}, N.~C. 2012, \aap, 545, A109,
  \dodoi{10.1051/0004-6361/201219115}

\bibitem[{{Brady} \& {Bean}(2022)}]{2022AJ....163..255B}
{Brady}, M.~T., \& {Bean}, J.~L. 2022, \aj, 163, 255,
  \dodoi{10.3847/1538-3881/ac64a0}

\bibitem[{{Burn} {et~al.}(2021){Burn}, {Schlecker}, {Mordasini}, {Emsenhuber},
  {Alibert}, {Henning}, {Klahr}, \& {Benz}}]{2021A&A...656A..72B}
{Burn}, R., {Schlecker}, M., {Mordasini}, C., {et~al.} 2021, \aap, 656, A72,
  \dodoi{10.1051/0004-6361/202140390}

\bibitem[{{Claret} {et~al.}(2013){Claret}, {Hauschildt}, \&
  {Witte}}]{2013A&A...552A..16C}
{Claret}, A., {Hauschildt}, P.~H., \& {Witte}, S. 2013, \aap, 552, A16,
  \dodoi{10.1051/0004-6361/201220942}

\bibitem[{{Crossfield} {et~al.}(2019){Crossfield}, {Waalkes}, {Newton},
  {Narita}, {Muirhead}, {Ment}, {Matthews}, {Kraus}, {Kostov}, {Kosiarek},
  {Kane}, {Isaacson}, {Halverson}, {Gonzales}, {Everett}, {Dragomir},
  {Collins}, {Chontos}, {Berardo}, {Winters}, {Winn}, {Scott}, {Rojas-Ayala},
  {Rizzuto}, {Petigura}, {Peterson}, {Mocnik}, {Mikal-Evans}, {Mehrle},
  {Matson}, {Kuzuhara}, {Irwin}, {Huber}, {Huang}, {Howell}, {Howard},
  {Hirano}, {Fulton}, {Dupuy}, {Dressing}, {Dalba}, {Charbonneau}, {Burt},
  {Berta-Thompson}, {Benneke}, {Watanabe}, {Twicken}, {Tamura}, {Schlieder},
  {Seager}, {Rose}, {Ricker}, {Quintana}, {L{\'e}pine}, {Latham}, {Kotani},
  {Jenkins}, {Hori}, {Colon}, \& {Caldwell}}]{2019ApJ...883L..16C}
{Crossfield}, I. J.~M., {Waalkes}, W., {Newton}, E.~R., {et~al.} 2019, \apjl,
  883, L16, \dodoi{10.3847/2041-8213/ab3d30}

\bibitem[{{Dawson} {et~al.}(2015){Dawson}, {Chiang}, \&
  {Lee}}]{2015MNRAS.453.1471D}
{Dawson}, R.~I., {Chiang}, E., \& {Lee}, E.~J. 2015, \mnras, 453, 1471,
  \dodoi{10.1093/mnras/stv1639}

\bibitem[{{Desort} {et~al.}(2007){Desort}, {Lagrange}, {Galland}, {Udry}, \&
  {Mayor}}]{2007A&A...473..983D}
{Desort}, M., {Lagrange}, A.~M., {Galland}, F., {Udry}, S., \& {Mayor}, M.
  2007, \aap, 473, 983, \dodoi{10.1051/0004-6361:20078144}

\bibitem[{{Dittmann} {et~al.}(2017){Dittmann}, {Irwin}, {Charbonneau},
  {Bonfils}, {Astudillo-Defru}, {Haywood}, {Berta-Thompson}, {Newton},
  {Rodriguez}, {Winters}, {Tan}, {Almenara}, {Bouchy}, {Delfosse}, {Forveille},
  {Lovis}, {Murgas}, {Pepe}, {Santos}, {Udry}, {W{\"u}nsche}, {Esquerdo},
  {Latham}, \& {Dressing}}]{2017Natur.544..333D}
{Dittmann}, J.~A., {Irwin}, J.~M., {Charbonneau}, D., {et~al.} 2017, \nat, 544,
  333, \dodoi{10.1038/nature22055}

\bibitem[{{Donati} {et~al.}(1997){Donati}, {Semel}, {Carter}, {Rees}, \&
  {Collier Cameron}}]{1997MNRAS.291..658D}
{Donati}, J.-F., {Semel}, M., {Carter}, B.~D., {Rees}, D.~E., \& {Collier
  Cameron}, A. 1997, \mnras, 291, 658, \dodoi{10.1093/mnras/291.4.658}

\bibitem[{{Dressing} \& {Charbonneau}(2015)}]{2015ApJ...807...45D}
{Dressing}, C.~D., \& {Charbonneau}, D. 2015, \apj, 807, 45,
  \dodoi{10.1088/0004-637X/807/1/45}

\bibitem[{{Dressing} {et~al.}(2017){Dressing}, {Vanderburg}, {Schlieder},
  {Crossfield}, {Knutson}, {Newton}, {Ciardi}, {Fulton}, {Gonzales}, {Howard},
  {Isaacson}, {Livingston}, {Petigura}, {Sinukoff}, {Everett}, {Horch}, \&
  {Howell}}]{2017AJ....154..207D}
{Dressing}, C.~D., {Vanderburg}, A., {Schlieder}, J.~E., {et~al.} 2017, \aj,
  154, 207, \dodoi{10.3847/1538-3881/aa89f2}

\bibitem[{{Elkins-Tanton} \& {Seager}(2008)}]{2008ApJ...685.1237E}
{Elkins-Tanton}, L.~T., \& {Seager}, S. 2008, \apj, 685, 1237,
  \dodoi{10.1086/591433}

\bibitem[{{Fridlund} {et~al.}(2017){Fridlund}, {Gaidos}, {Barrag{\'a}n},
  {Persson}, {Gandolfi}, {Cabrera}, {Hirano}, {Kuzuhara}, {Csizmadia}, {Nowak},
  {Endl}, {Grziwa}, {Korth}, {Pfaff}, {Bitsch}, {Johansen}, {Mustill},
  {Davies}, {Deeg}, {Palle}, {Cochran}, {Eigm{\"u}ller}, {Erikson}, {Guenther},
  {Hatzes}, {Kiilerich}, {Kudo}, {MacQueen}, {Narita}, {Nespral},
  {P{\"a}tzold}, {Prieto-Arranz}, {Rauer}, \& {Van
  Eylen}}]{2017A&A...604A..16F}
{Fridlund}, M., {Gaidos}, E., {Barrag{\'a}n}, O., {et~al.} 2017, \aap, 604,
  A16, \dodoi{10.1051/0004-6361/201730822}

\bibitem[{{Gaia Collaboration} {et~al.}(2021){Gaia Collaboration}, {Brown},
  {Vallenari}, {Prusti}, {de Bruijne}, {Babusiaux}, {Biermann}, {Creevey},
  {Evans}, {Eyer}, {Hutton}, {Jansen}, {Jordi}, {Klioner}, {Lammers},
  {Lindegren}, {Luri}, {Mignard}, {Panem}, {Pourbaix}, {Randich}, {Sartoretti},
  {Soubiran}, {Walton}, {Arenou}, {Bailer-Jones}, {Bastian}, {Cropper},
  {Drimmel}, {Katz}, {Lattanzi}, {van Leeuwen}, {Bakker}, {Cacciari},
  {Casta{\~n}eda}, {De Angeli}, {Ducourant}, {Fabricius}, {Fouesneau},
  {Fr{\'e}mat}, {Guerra}, {Guerrier}, {Guiraud}, {Jean-Antoine Piccolo},
  {Masana}, {Messineo}, {Mowlavi}, {Nicolas}, {Nienartowicz}, {Pailler},
  {Panuzzo}, {Riclet}, {Roux}, {Seabroke}, {Sordo}, {Tanga}, {Th{\'e}venin},
  {Gracia-Abril}, {Portell}, {Teyssier}, {Altmann}, {Andrae}, {Bellas-Velidis},
  {Benson}, {Berthier}, {Blomme}, {Brugaletta}, {Burgess}, {Busso}, {Carry},
  {Cellino}, {Cheek}, {Clementini}, {Damerdji}, {Davidson}, {Delchambre},
  {Dell'Oro}, {Fern{\'a}ndez-Hern{\'a}ndez}, {Galluccio}, {Garc{\'\i}a-Lario},
  {Garcia-Reinaldos}, {Gonz{\'a}lez-N{\'u}{\~n}ez}, {Gosset}, {Haigron},
  {Halbwachs}, {Hambly}, {Harrison}, {Hatzidimitriou}, {Heiter},
  {Hern{\'a}ndez}, {Hestroffer}, {Hodgkin}, {Holl}, {Jan{\ss}en}, {Jevardat de
  Fombelle}, {Jordan}, {Krone-Martins}, {Lanzafame}, {L{\"o}ffler}, {Lorca},
  {Manteiga}, {Marchal}, {Marrese}, {Moitinho}, {Mora}, {Muinonen}, {Osborne},
  {Pancino}, {Pauwels}, {Petit}, {Recio-Blanco}, {Richards}, {Riello},
  {Rimoldini}, {Robin}, {Roegiers}, {Rybizki}, {Sarro}, {Siopis}, {Smith},
  {Sozzetti}, {Ulla}, {Utrilla}, {van Leeuwen}, {van Reeven}, {Abbas}, {Abreu
  Aramburu}, {Accart}, {Aerts}, {Aguado}, {Ajaj}, {Altavilla}, {{\'A}lvarez},
  {{\'A}lvarez Cid-Fuentes}, {Alves}, {Anderson}, {Anglada Varela}, {Antoja},
  {Audard}, {Baines}, {Baker}, {Balaguer-N{\'u}{\~n}ez}, {Balbinot}, {Balog},
  {Barache}, {Barbato}, {Barros}, {Barstow}, {Bartolom{\'e}}, {Bassilana},
  {Bauchet}, {Baudesson-Stella}, {Becciani}, {Bellazzini}, {Bernet}, {Bertone},
  {Bianchi}, {Blanco-Cuaresma}, {Boch}, {Bombrun}, {Bossini}, {Bouquillon},
  {Bragaglia}, {Bramante}, {Breedt}, {Bressan}, {Brouillet}, {Bucciarelli},
  {Burlacu}, {Busonero}, {Butkevich}, {Buzzi}, {Caffau}, {Cancelliere},
  {C{\'a}novas}, {Cantat-Gaudin}, {Carballo}, {Carlucci}, {Carnerero},
  {Carrasco}, {Casamiquela}, {Castellani}, {Castro-Ginard}, {Castro Sampol},
  {Chaoul}, {Charlot}, {Chemin}, {Chiavassa}, {Cioni}, {Comoretto}, {Cooper},
  {Cornez}, {Cowell}, {Crifo}, {Crosta}, {Crowley}, {Dafonte}, {Dapergolas},
  {David}, {David}, {de Laverny}, {De Luise}, {De March}, {De Ridder}, {de
  Souza}, {de Teodoro}, {de Torres}, {del Peloso}, {del Pozo}, {Delbo},
  {Delgado}, {Delgado}, {Delisle}, {Di Matteo}, {Diakite}, {Diener},
  {Distefano}, {Dolding}, {Eappachen}, {Edvardsson}, {Enke}, {Esquej}, {Fabre},
  {Fabrizio}, {Faigler}, {Fedorets}, {Fernique}, {Fienga}, {Figueras},
  {Fouron}, {Fragkoudi}, {Fraile}, {Franke}, {Gai}, {Garabato},
  {Garcia-Gutierrez}, {Garc{\'\i}a-Torres}, {Garofalo}, {Gavras}, {Gerlach},
  {Geyer}, {Giacobbe}, {Gilmore}, {Girona}, {Giuffrida}, {Gomel}, {Gomez},
  {Gonzalez-Santamaria}, {Gonz{\'a}lez-Vidal}, {Granvik},
  {Guti{\'e}rrez-S{\'a}nchez}, {Guy}, {Hauser}, {Haywood}, {Helmi}, {Hidalgo},
  {Hilger}, {H{\l}adczuk}, {Hobbs}, {Holland}, {Huckle}, {Jasniewicz},
  {Jonker}, {Juaristi Campillo}, {Julbe}, {Karbevska}, {Kervella}, {Khanna},
  {Kochoska}, {Kontizas}, {Kordopatis}, {Korn}, {Kostrzewa-Rutkowska},
  {Kruszy{\'n}ska}, {Lambert}, {Lanza}, {Lasne}, {Le Campion}, {Le Fustec},
  {Lebreton}, {Lebzelter}, {Leccia}, {Leclerc}, {Lecoeur-Taibi}, {Liao},
  {Licata}, {Lindstr{\o}m}, {Lister}, {Livanou}, {Lobel}, {Madrero Pardo},
  {Managau}, {Mann}, {Marchant}, {Marconi}, {Marcos Santos}, {Marinoni},
  {Marocco}, {Marshall}, {Martin Polo}, {Mart{\'\i}n-Fleitas}, {Masip},
  {Massari}, {Mastrobuono-Battisti}, {Mazeh}, {McMillan}, {Messina},
  {Michalik}, {Millar}, {Mints}, {Molina}, {Molinaro}, {Moln{\'a}r},
  {Montegriffo}, {Mor}, {Morbidelli}, {Morel}, {Morris}, {Mulone}, {Munoz},
  {Muraveva}, {Murphy}, {Musella}, {Noval}, {Ord{\'e}novic}, {Orr{\`u}},
  {Osinde}, {Pagani}, {Pagano}, {Palaversa}, {Palicio}, {Panahi}, {Pawlak},
  {Pe{\~n}alosa Esteller}, {Penttil{\"a}}, {Piersimoni}, {Pineau}, {Plachy},
  {Plum}, {Poggio}, {Poretti}, {Poujoulet}, {Pr{\v{s}}a}, {Pulone}, {Racero},
  {Ragaini}, {Rainer}, {Raiteri}, {Rambaux}, {Ramos}, {Ramos-Lerate}, {Re
  Fiorentin}, {Regibo}, {Reyl{\'e}}, {Ripepi}, {Riva}, {Rixon}, {Robichon},
  {Robin}, {Roelens}, {Rohrbasser}, {Romero-G{\'o}mez}, {Rowell}, {Royer},
  {Rybicki}, {Sadowski}, {Sagrist{\`a} Sell{\'e}s}, {Sahlmann}, {Salgado},
  {Salguero}, {Samaras}, {Sanchez Gimenez}, {Sanna}, {Santove{\~n}a},
  {Sarasso}, {Schultheis}, {Sciacca}, {Segol}, {Segovia}, {S{\'e}gransan},
  {Semeux}, {Shahaf}, {Siddiqui}, {Siebert}, {Siltala}, {Slezak}, {Smart},
  {Solano}, {Solitro}, {Souami}, {Souchay}, {Spagna}, {Spoto}, {Steele},
  {Steidelm{\"u}ller}, {Stephenson}, {S{\"u}veges}, {Szabados}, {Szegedi-Elek},
  {Taris}, {Tauran}, {Taylor}, {Teixeira}, {Thuillot}, {Tonello}, {Torra},
  {Torra}, {Turon}, {Unger}, {Vaillant}, {van Dillen}, {Vanel}, {Vecchiato},
  {Viala}, {Vicente}, {Voutsinas}, {Weiler}, {Wevers}, {Wyrzykowski}, {Yoldas},
  {Yvard}, {Zhao}, {Zorec}, {Zucker}, {Zurbach}, \&
  {Zwitter}}]{2021A&A...649A...1G}
{Gaia Collaboration}, {Brown}, A.~G.~A., {Vallenari}, A., {et~al.} 2021, \aap,
  649, A1, \dodoi{10.1051/0004-6361/202039657}

\bibitem[{{Gaidos} {et~al.}(2016){Gaidos}, {Mann}, {Kraus}, \&
  {Ireland}}]{2016MNRAS.457.2877G}
{Gaidos}, E., {Mann}, A.~W., {Kraus}, A.~L., \& {Ireland}, M. 2016, \mnras,
  457, 2877, \dodoi{10.1093/mnras/stw097}

\bibitem[{{Gandolfi} {et~al.}(2017){Gandolfi}, {Barrag{\'a}n}, {Hatzes},
  {Fridlund}, {Fossati}, {Donati}, {Johnson}, {Nowak}, {Prieto-Arranz},
  {Albrecht}, {Dai}, {Deeg}, {Endl}, {Grziwa}, {Hjorth}, {Korth}, {Nespral},
  {Saario}, {Smith}, {Antoniciello}, {Alarcon}, {Bedell}, {Blay}, {Brems},
  {Cabrera}, {Csizmadia}, {Cusano}, {Cochran}, {Eigm{\"u}ller}, {Erikson},
  {Gonz{\'a}lez Hern{\'a}ndez}, {Guenther}, {Hirano}, {Su{\'a}rez
  Mascare{\~n}o}, {Narita}, {Palle}, {Parviainen}, {P{\"a}tzold}, {Persson},
  {Rauer}, {Saviane}, {Schmidtobreick}, {Van Eylen}, {Winn}, \&
  {Zakhozhay}}]{2017AJ....154..123G}
{Gandolfi}, D., {Barrag{\'a}n}, O., {Hatzes}, A.~P., {et~al.} 2017, \aj, 154,
  123, \dodoi{10.3847/1538-3881/aa832a}

\bibitem[{{Gardner} {et~al.}(2006){Gardner}, {Mather}, {Clampin}, {Doyon},
  {Greenhouse}, {Hammel}, {Hutchings}, {Jakobsen}, {Lilly}, {Long}, {Lunine},
  {McCaughrean}, {Mountain}, {Nella}, {Rieke}, {Rieke}, {Rix}, {Smith},
  {Sonneborn}, {Stiavelli}, {Stockman}, {Windhorst}, \&
  {Wright}}]{2006SSRv..123..485G}
{Gardner}, J.~P., {Mather}, J.~C., {Clampin}, M., {et~al.} 2006, \ssr, 123,
  485, \dodoi{10.1007/s11214-006-8315-7}

\bibitem[{{Giacalone} {et~al.}(2021){Giacalone}, {Dressing}, {Jensen},
  {Collins}, {Ricker}, {Vanderspek}, {Seager}, {Winn}, {Jenkins}, {Barclay},
  {Barkaoui}, {Cadieux}, {Charbonneau}, {Collins}, {Conti}, {Doyon}, {Evans},
  {Ghachoui}, {Gillon}, {Guerrero}, {Hart}, {Jehin}, {Kielkopf}, {McLean},
  {Murgas}, {Palle}, {Parviainen}, {Pozuelos}, {Relles}, {Shporer}, {Socia},
  {Stockdale}, {Tan}, {Torres}, {Twicken}, {Waalkes}, \&
  {Waite}}]{2021AJ....161...24G}
{Giacalone}, S., {Dressing}, C.~D., {Jensen}, E. L.~N., {et~al.} 2021, \aj,
  161, 24, \dodoi{10.3847/1538-3881/abc6af}

\bibitem[{{Gillon} {et~al.}(2017){Gillon}, {Triaud}, {Demory}, {Jehin}, {Agol},
  {Deck}, {Lederer}, {de Wit}, {Burdanov}, {Ingalls}, {Bolmont}, {Leconte},
  {Raymond}, {Selsis}, {Turbet}, {Barkaoui}, {Burgasser}, {Burleigh}, {Carey},
  {Chaushev}, {Copperwheat}, {Delrez}, {Fernandes}, {Holdsworth}, {Kotze}, {Van
  Grootel}, {Almleaky}, {Benkhaldoun}, {Magain}, \&
  {Queloz}}]{2017Natur.542..456G}
{Gillon}, M., {Triaud}, A.~H.~M.~J., {Demory}, B.-O., {et~al.} 2017, \nat, 542,
  456, \dodoi{10.1038/nature21360}

\bibitem[{{Girardi} {et~al.}(2005){Girardi}, {Groenewegen}, {Hatziminaoglou},
  \& {da Costa}}]{Girardi2005}
{Girardi}, L., {Groenewegen}, M.~A.~T., {Hatziminaoglou}, E., \& {da Costa}, L.
  2005, \aap, 436, 895, \dodoi{10.1051/0004-6361:20042352}

\bibitem[{{Grunblatt} {et~al.}(2015){Grunblatt}, {Howard}, \&
  {Haywood}}]{2015ApJ...808..127G}
{Grunblatt}, S.~K., {Howard}, A.~W., \& {Haywood}, R.~D. 2015, \apj, 808, 127,
  \dodoi{10.1088/0004-637X/808/2/127}

\bibitem[{{Guerrero} {et~al.}(2021){Guerrero}, {Seager}, {Huang}, {Vanderburg},
  {Garcia Soto}, {Mireles}, {Hesse}, {Fong}, {Glidden}, {Shporer}, {Latham},
  {Collins}, {Quinn}, {Burt}, {Dragomir}, {Crossfield}, {Vanderspek},
  {Fausnaugh}, {Burke}, {Ricker}, {Daylan}, {Essack}, {G{\"u}nther}, {Osborn},
  {Pepper}, {Rowden}, {Sha}, {Villanueva}, {Yahalomi}, {Yu}, {Ballard},
  {Batalha}, {Berardo}, {Chontos}, {Dittmann}, {Esquerdo}, {Mikal-Evans},
  {Jayaraman}, {Krishnamurthy}, {Louie}, {Mehrle}, {Niraula}, {Rackham},
  {Rodriguez}, {Rowden}, {Sousa-Silva}, {Watanabe}, {Wong}, {Zhan},
  {Zivanovic}, {Christiansen}, {Ciardi}, {Swain}, {Lund}, {Mullally},
  {Fleming}, {Rodriguez}, {Boyd}, {Quintana}, {Barclay}, {Col{\'o}n},
  {Rinehart}, {Schlieder}, {Clampin}, {Jenkins}, {Twicken}, {Caldwell},
  {Coughlin}, {Henze}, {Lissauer}, {Morris}, {Rose}, {Smith}, {Tenenbaum},
  {Ting}, {Wohler}, {Bakos}, {Bean}, {Berta-Thompson}, {Bieryla}, {Bouma},
  {Buchhave}, {Butler}, {Charbonneau}, {Doty}, {Ge}, {Holman}, {Howard},
  {Kaltenegger}, {Kane}, {Kjeldsen}, {Kreidberg}, {Lin}, {Minsky}, {Narita},
  {Paegert}, {P{\'a}l}, {Palle}, {Sasselov}, {Spencer}, {Sozzetti}, {Stassun},
  {Torres}, {Udry}, \& {Winn}}]{2021ApJS..254...39G}
{Guerrero}, N.~M., {Seager}, S., {Huang}, C.~X., {et~al.} 2021, \apjs, 254, 39,
  \dodoi{10.3847/1538-4365/abefe1}

\bibitem[{{Harakawa} {et~al.}(2022){Harakawa}, {Takarada}, {Kasagi}, {Hirano},
  {Kotani}, {Kuzuhara}, {Omiya}, {Kawahara}, {Fukui}, {Hori}, {Ishikawa},
  {Ogihara}, {Livingston}, {Brandt}, {Currie}, {Aoki}, {Beichman}, {Henning},
  {Hodapp}, {Ishizuka}, {Izumiura}, {Jacobson}, {Janson}, {Kambe}, {Kodama},
  {Kokubo}, {Konishi}, {Krishnamurthy}, {Kudo}, {Kurokawa}, {Kusakabe}, {Kwon},
  {Matsumoto}, {McElwain}, {Mitsui}, {Nakagawa}, {Narita}, {Nishikawa},
  {Nugroho}, {Serabyn}, {Serizawa}, {Takahashi}, {Ueda}, {Uyama}, {Vievard},
  {Wang}, {Wisniewski}, {Tamura}, \& {Sato}}]{2022PASJ...74..904H}
{Harakawa}, H., {Takarada}, T., {Kasagi}, Y., {et~al.} 2022, \pasj, 74, 904,
  \dodoi{10.1093/pasj/psac044}

\bibitem[{{Hardegree-Ullman} {et~al.}(2019){Hardegree-Ullman}, {Cushing},
  {Muirhead}, \& {Christiansen}}]{2019AJ....158...75H}
{Hardegree-Ullman}, K.~K., {Cushing}, M.~C., {Muirhead}, P.~S., \&
  {Christiansen}, J.~L. 2019, \aj, 158, 75, \dodoi{10.3847/1538-3881/ab21d2}

\bibitem[{{Hayano} {et~al.}(2008){Hayano}, {Takami}, {Guyon}, {Oya}, {Hattori},
  {Saito}, {Watanabe}, {Murakami}, {Minowa}, {Ito}, {Colley}, {Eldred},
  {Golota}, {Dinkins}, {Kashikawa}, \& {Iye}}]{2008SPIE.7015E..10H}
{Hayano}, Y., {Takami}, H., {Guyon}, O., {et~al.} 2008, \procspie, Vol. 7015,
  {Current status of the laser guide star adaptive optics system for Subaru
  Telescope} (SPIE), 701510, \dodoi{10.1117/12.789992}

\bibitem[{{Henden} {et~al.}(2016){Henden}, {Templeton}, {Terrell}, {Smith},
  {Levine}, \& {Welch}}]{2016yCat.2336....0H}
{Henden}, A.~A., {Templeton}, M., {Terrell}, D., {et~al.} 2016, VizieR Online
  Data Catalog, 2336

\bibitem[{{Hirano} {et~al.}(2015){Hirano}, {Masuda}, {Sato}, {Benomar},
  {Takeda}, {Omiya}, {Harakawa}, \& {Kobayashi}}]{2015ApJ...799....9H}
{Hirano}, T., {Masuda}, K., {Sato}, B., {et~al.} 2015, \apj, 799, 9,
  \dodoi{10.1088/0004-637X/799/1/9}

\bibitem[{{Hirano} {et~al.}(2016){Hirano}, {Fukui}, {Mann}, {Sanchis-Ojeda},
  {Gaidos}, {Narita}, {Dai}, {Van Eylen}, {Lee}, {Onozato}, {Ryu}, {Kusakabe},
  {Ito}, {Kuzuhara}, {Onitsuka}, {Tatsuuma}, {Nowak}, {Pall{\`e}}, {Ribas},
  {Tamura}, \& {Yu}}]{2016ApJ...820...41H}
{Hirano}, T., {Fukui}, A., {Mann}, A.~W., {et~al.} 2016, \apj, 820, 41,
  \dodoi{10.3847/0004-637X/820/1/41}

\bibitem[{{Hirano} {et~al.}(2018){Hirano}, {Dai}, {Gandolfi}, {Fukui},
  {Livingston}, {Miyakawa}, {Endl}, {Cochran}, {Alonso-Floriano}, {Kuzuhara},
  {Montes}, {Ryu}, {Albrecht}, {Barragan}, {Cabrera}, {Csizmadia}, {Deeg},
  {Eigm{\"u}ller}, {Erikson}, {Fridlund}, {Grziwa}, {Guenther}, {Hatzes},
  {Korth}, {Kudo}, {Kusakabe}, {Narita}, {Nespral}, {Nowak}, {P{\"a}tzold},
  {Palle}, {Persson}, {Prieto-Arranz}, {Rauer}, {Ribas}, {Sato}, {Smith},
  {Tamura}, {Tanaka}, {Van Eylen}, \& {Winn}}]{2018AJ....155..127H}
{Hirano}, T., {Dai}, F., {Gandolfi}, D., {et~al.} 2018, \aj, 155, 127,
  \dodoi{10.3847/1538-3881/aaa9c1}

\bibitem[{{Hirano} {et~al.}(2020){Hirano}, {Kuzuhara}, {Kotani}, {Omiya},
  {Kudo}, {Harakawa}, {Vievard}, {Kurokawa}, {Nishikawa}, {Tamura}, {Hodapp},
  {Ishizuka}, {Jacobson}, {Konishi}, {Serizawa}, {Ueda}, {Gaidos}, \&
  {Sato}}]{2020PASJ...72...93H}
{Hirano}, T., {Kuzuhara}, M., {Kotani}, T., {et~al.} 2020, \pasj, 72, 93,
  \dodoi{10.1093/pasj/psaa085}

\bibitem[{{Hirano} {et~al.}(2021){Hirano}, {Livingston}, {Fukui}, {Narita},
  {Harakawa}, {Ishikawa}, {Miyakawa}, {Kimura}, {Nakayama}, {Fujita}, {Hori},
  {Stassun}, {Bieryla}, {Cadieux}, {Ciardi}, {Collins}, {Ikoma}, {Vanderburg},
  {Barclay}, {Brasseur}, {de Leon}, {Doty}, {Doyon}, {Esparza-Borges},
  {Esquerdo}, {Furlan}, {Gaidos}, {Gonzales}, {Hodapp}, {Howell}, {Isogai},
  {Jacobson}, {Jenkins}, {Jensen}, {Kawauchi}, {Kotani}, {Kudo}, {Kurita},
  {Kurokawa}, {Kusakabe}, {Kuzuhara}, {Lafreni{\`e}re}, {Latham}, {Massey},
  {Mori}, {Murgas}, {Nishikawa}, {Nishiumi}, {Omiya}, {Paegert}, {Palle},
  {Parviainen}, {Quinn}, {Ricker}, {Schwarz}, {Seager}, {Tamura}, {Tenenbaum},
  {Terada}, {Vanderspek}, {Vievard}, {Watanabe}, \&
  {Winn}}]{2021AJ....162..161H}
{Hirano}, T., {Livingston}, J.~H., {Fukui}, A., {et~al.} 2021, \aj, 162, 161,
  \dodoi{10.3847/1538-3881/ac0fdc}

\bibitem[{{Hoshino} \& {Kokubo}(2023)}]{2023MNRAS.519.2838H}
{Hoshino}, H., \& {Kokubo}, E. 2023, \mnras, 519, 2838,
  \dodoi{10.1093/mnras/stac3756}

\bibitem[{{Howell} {et~al.}(2011){Howell}, {Everett}, {Sherry}, {Horch}, \&
  {Ciardi}}]{Howell2011}
{Howell}, S.~B., {Everett}, M.~E., {Sherry}, W., {Horch}, E., \& {Ciardi},
  D.~R. 2011, \aj, 142, 19, \dodoi{10.1088/0004-6256/142/1/19}

\bibitem[{{Howell} {et~al.}(2014){Howell}, {Sobeck}, {Haas}, {Still},
  {Barclay}, {Mullally}, {Troeltzsch}, {Aigrain}, {Bryson}, {Caldwell},
  {Chaplin}, {Cochran}, {Huber}, {Marcy}, {Miglio}, {Najita}, {Smith},
  {Twicken}, \& {Fortney}}]{Howell2014}
{Howell}, S.~B., {Sobeck}, C., {Haas}, M., {et~al.} 2014, \pasp, 126, 398,
  \dodoi{10.1086/676406}

\bibitem[{{Ikoma} \& {Genda}(2006)}]{2006ApJ...648..696I}
{Ikoma}, M., \& {Genda}, H. 2006, \apj, 648, 696, \dodoi{10.1086/505780}

\bibitem[{{Ishikawa} {et~al.}(2020){Ishikawa}, {Aoki}, {Kotani}, {Kuzuhara},
  {Omiya}, {Reiners}, \& {Zechmeister}}]{2020PASJ...72..102I}
{Ishikawa}, H.~T., {Aoki}, W., {Kotani}, T., {et~al.} 2020, \pasj, 72, 102,
  \dodoi{10.1093/pasj/psaa101}

\bibitem[{{Ishikawa} {et~al.}(2022){Ishikawa}, {Aoki}, {Hirano}, {Kotani},
  {Kuzuhara}, {Omiya}, {Hori}, {Kokubo}, {Kudo}, {Kurokawa}, {Kusakabe},
  {Narita}, {Nishikawa}, {Ogihara}, {Ueda}, {Currie}, {Henning}, {Kasagi},
  {Kolecki}, {Kwon}, {Machida}, {McElwain}, {Nakagawa}, {Vievard}, {Wang},
  {Tamura}, \& {Sato}}]{2022AJ....163...72I}
{Ishikawa}, H.~T., {Aoki}, W., {Hirano}, T., {et~al.} 2022, \aj, 163, 72,
  \dodoi{10.3847/1538-3881/ac3ee0}

\bibitem[{{Johnstone}(2020)}]{2020ApJ...890...79J}
{Johnstone}, C.~P. 2020, \apj, 890, 79, \dodoi{10.3847/1538-4357/ab6224}

\bibitem[{{Kempton} {et~al.}(2018){Kempton}, {Bean}, {Louie}, {Deming}, {Koll},
  {Mansfield}, {Christiansen}, {L{\'o}pez-Morales}, {Swain}, {Zellem},
  {Ballard}, {Barclay}, {Barstow}, {Batalha}, {Beatty}, {Berta-Thompson},
  {Birkby}, {Buchhave}, {Charbonneau}, {Cowan}, {Crossfield}, {de Val-Borro},
  {Doyon}, {Dragomir}, {Gaidos}, {Heng}, {Hu}, {Kane}, {Kreidberg}, {Mallonn},
  {Morley}, {Narita}, {Nascimbeni}, {Pall{\'e}}, {Quintana}, {Rauscher},
  {Seager}, {Shkolnik}, {Sing}, {Sozzetti}, {Stassun}, {Valenti}, \& {von
  Essen}}]{2018PASP..130k4401K}
{Kempton}, E. M.~R., {Bean}, J.~L., {Louie}, D.~R., {et~al.} 2018, \pasp, 130,
  114401, \dodoi{10.1088/1538-3873/aadf6f}

\bibitem[{{Kimura} \& {Ikoma}(2022)}]{Kimura+2022}
{Kimura}, T., \& {Ikoma}, M. 2022, Nature Astronomy,
  \dodoi{10.1038/s41550-022-01781-1}

\bibitem[{{Kite} \& {Schaefer}(2021)}]{2021ApJ...909L..22K}
{Kite}, E.~S., \& {Schaefer}, L. 2021, \apjl, 909, L22,
  \dodoi{10.3847/2041-8213/abe7dc}

\bibitem[{{Kobayashi} {et~al.}(2000){Kobayashi}, {Tokunaga}, {Terada}, {Goto},
  {Weber}, {Potter}, {Onaka}, {Ching}, {Young}, {Fletcher}, {Neil},
  {Robertson}, {Cook}, {Imanishi}, \& {Warren}}]{2000SPIE.4008.1056K}
{Kobayashi}, N., {Tokunaga}, A.~T., {Terada}, H., {et~al.} 2000, in \procspie,
  Vol. 4008, Optical and IR Telescope Instrumentation and Detectors, ed.
  M.~{Iye} \& A.~F. {Moorwood}, 1056--1066, \dodoi{10.1117/12.395423}

\bibitem[{{Koizumi} {et~al.}(2021){Koizumi}, {Kuzuhara}, {Omiya}, {Hirano},
  {Wisniewski}, {Aoki}, \& {Sato}}]{2021PASJ...73..154K}
{Koizumi}, Y., {Kuzuhara}, M., {Omiya}, M., {et~al.} 2021, \pasj, 73, 154,
  \dodoi{10.1093/pasj/psaa112}

\bibitem[{{Kopparapu} {et~al.}(2016){Kopparapu}, {Wolf}, {Haqq-Misra}, {Yang},
  {Kasting}, {Meadows}, {Terrien}, \& {Mahadevan}}]{2016ApJ...819...84K}
{Kopparapu}, R.~k., {Wolf}, E.~T., {Haqq-Misra}, J., {et~al.} 2016, \apj, 819,
  84, \dodoi{10.3847/0004-637X/819/1/84}

\bibitem[{{Kotani} {et~al.}(2018){Kotani}, {Tamura}, {Nishikawa}, {Ueda},
  {Kuzuhara}, {Omiya}, {Hashimoto}, {Ishizuka}, {Hirano}, {Suto}, {Kurokawa},
  {Kokubo}, {Mori}, {Tanaka}, {Kashiwagi}, {Konishi}, {Kudo}, {Sato},
  {Jacobson}, {Hodapp}, {Hall}, {Aoki}, {Usuda}, {Nishiyama}, {Nakajima},
  {Ikeda}, {Yamamuro}, {Morino}, {Baba}, {Hosokawa}, {Ishikawa}, {Narita},
  {Kokubo}, {Hayano}, {Izumiura}, {Kambe}, {Kusakabe}, {Kwon}, {Ikoma}, {Hori},
  {Genda}, {Fukui}, {Fujii}, {Kawahara}, {Olivier}, {Jovanovic}, {Harakawa},
  {Hayashi}, {Hidai}, {Machida}, {Matsuo}, {Nagata}, {Ogihara}, {Takami},
  {Takato}, {Terada}, \& {Oh}}]{2018SPIE10702E..11K}
{Kotani}, T., {Tamura}, M., {Nishikawa}, J., {et~al.} 2018, in \procspie, Vol.
  10702, Ground-based and Airborne Instrumentation for Astronomy VII, 1070211,
  \dodoi{10.1117/12.2311836}

\bibitem[{{Kov{\'a}cs} {et~al.}(2002){Kov{\'a}cs}, {Zucker}, \&
  {Mazeh}}]{Kovacs}
{Kov{\'a}cs}, G., {Zucker}, S., \& {Mazeh}, T. 2002, \aap, 391, 369,
  \dodoi{10.1051/0004-6361:20020802}

\bibitem[{{Kreidberg}(2018)}]{2018haex.bookE.100K}
{Kreidberg}, L. 2018, in Handbook of Exoplanets, ed. H.~J. {Deeg} \& J.~A.
  {Belmonte}, 100, \dodoi{10.1007/978-3-319-55333-7_100}

\bibitem[{{Lammer} {et~al.}(2007){Lammer}, {Lichtenegger}, {Kulikov},
  {Grie{\ss}meier}, {Terada}, {Erkaev}, {Biernat}, {Khodachenko}, {Ribas},
  {Penz}, \& {Selsis}}]{2007AsBio...7..185L}
{Lammer}, H., {Lichtenegger}, H. I.~M., {Kulikov}, Y.~N., {et~al.} 2007,
  Astrobiology, 7, 185, \dodoi{10.1089/ast.2006.0128}

\bibitem[{{L{\'e}pine} \& {Shara}(2005)}]{2005AJ....129.1483L}
{L{\'e}pine}, S., \& {Shara}, M.~M. 2005, \aj, 129, 1483,
  \dodoi{10.1086/427854}

\bibitem[{{Lichtenberg} {et~al.}(2021){Lichtenberg}, {Bower}, {Hammond},
  {Boukrouche}, {Sanan}, {Tsai}, \& {Pierrehumbert}}]{2021JGRE..12606711L}
{Lichtenberg}, T., {Bower}, D.~J., {Hammond}, M., {et~al.} 2021, Journal of
  Geophysical Research (Planets), 126, e06711, \dodoi{10.1029/2020JE006711}

\bibitem[{{Liu} {et~al.}(2019){Liu}, {Lambrechts}, {Johansen}, \&
  {Liu}}]{2019A&A...632A...7L}
{Liu}, B., {Lambrechts}, M., {Johansen}, A., \& {Liu}, F. 2019, \aap, 632, A7,
  \dodoi{10.1051/0004-6361/201936309}

\bibitem[{{Liu} {et~al.}(2020){Liu}, {Lambrechts}, {Johansen}, {Pascucci}, \&
  {Henning}}]{2020A&A...638A..88L}
{Liu}, B., {Lambrechts}, M., {Johansen}, A., {Pascucci}, I., \& {Henning}, T.
  2020, \aap, 638, A88, \dodoi{10.1051/0004-6361/202037720}

\bibitem[{{Lopez} \& {Fortney}(2014)}]{2014ApJ...792....1L}
{Lopez}, E.~D., \& {Fortney}, J.~J. 2014, \apj, 792, 1,
  \dodoi{10.1088/0004-637X/792/1/1}

\bibitem[{{Luger} \& {Barnes}(2015)}]{2015AsBio..15..119L}
{Luger}, R., \& {Barnes}, R. 2015, Astrobiology, 15, 119,
  \dodoi{10.1089/ast.2014.1231}

\bibitem[{{Luo} {et~al.}(2015){Luo}, {Zhao}, {Zhao}, {Deng}, {Liu}, {Jing},
  {Wang}, {Zhang}, {Shi}, {Cui}, {Chu}, {Li}, {Bai}, {Wu}, {Cai}, {Cao}, {Cao},
  {Carlin}, {Chen}, {Chen}, {Chen}, {Chen}, {Chen}, {Chen}, {Chen},
  {Christlieb}, {Chu}, {Cui}, {Dong}, {Du}, {Fan}, {Feng}, {Fu}, {Gao}, {Gong},
  {Gu}, {Guo}, {Han}, {He}, {Hou}, {Hou}, {Hou}, {Hu}, {Hu}, {Hu}, {Huo},
  {Jia}, {Jiang}, {Jiang}, {Jiang}, {Jin}, {Kong}, {Kong}, {Lei}, {Li}, {Li},
  {Li}, {Li}, {Li}, {Li}, {Li}, {Li}, {Li}, {Li}, {Li}, {Li}, {Liang}, {Lin},
  {Liu}, {Liu}, {Liu}, {Liu}, {Lu}, {Luo}, {Mao}, {Newberg}, {Ni}, {Qi}, {Qi},
  {Shen}, {Shi}, {Song}, {Song}, {Su}, {Su}, {Tang}, {Tao}, {Tian}, {Wang},
  {Wang}, {Wang}, {Wang}, {Wang}, {Wang}, {Wang}, {Wang}, {Wang}, {Wang},
  {Wang}, {Wang}, {Wang}, {Wang}, {Wang}, {Wang}, {Wang}, {Wang}, {Wang},
  {Wang}, {Wei}, {Wei}, {Wu}, {Wu}, {Wu}, {Wu}, {Xing}, {Xu}, {Xu}, {Xu},
  {Yan}, {Yang}, {Yang}, {Yang}, {Yang}, {Yao}, {Yu}, {Yuan}, {Yuan}, {Yuan},
  {Yuan}, {Zhai}, {Zhang}, {Zhang}, {Zhang}, {Zhang}, {Zhang}, {Zhang},
  {Zhang}, {Zhang}, {Zhao}, {Zhou}, {Zhou}, {Zhu}, {Zhu}, {Zou}, \&
  {Zuo}}]{2015RAA....15.1095L}
{Luo}, A.~L., {Zhao}, Y.-H., {Zhao}, G., {et~al.} 2015, Research in Astronomy
  and Astrophysics, 15, 1095, \dodoi{10.1088/1674-4527/15/8/002}

\bibitem[{{Luque} \& {Pall{\'e}}(2022)}]{2022Sci...377.1211L}
{Luque}, R., \& {Pall{\'e}}, E. 2022, Science, 377, 1211,
  \dodoi{10.1126/science.abl7164}

\bibitem[{{Mann} {et~al.}(2015){Mann}, {Feiden}, {Gaidos}, {Boyajian}, \& {von
  Braun}}]{2015ApJ...804...64M}
{Mann}, A.~W., {Feiden}, G.~A., {Gaidos}, E., {Boyajian}, T., \& {von Braun},
  K. 2015, \apj, 804, 64, \dodoi{10.1088/0004-637X/804/1/64}

\bibitem[{{Mann} {et~al.}(2019){Mann}, {Dupuy}, {Kraus}, {Gaidos}, {Ansdell},
  {Ireland}, {Rizzuto}, {Hung}, {Dittmann}, {Factor}, {Feiden}, {Martinez},
  {Ru{\'\i}z-Rodr{\'\i}guez}, \& {Thao}}]{2019ApJ...871...63M}
{Mann}, A.~W., {Dupuy}, T., {Kraus}, A.~L., {et~al.} 2019, \apj, 871, 63,
  \dodoi{10.3847/1538-4357/aaf3bc}

\bibitem[{{Matsumoto} {et~al.}(2020){Matsumoto}, {Gu}, {Kokubo}, {Oshino}, \&
  {Omiya}}]{2020A&A...642A..23M}
{Matsumoto}, Y., {Gu}, P.-G., {Kokubo}, E., {Oshino}, S., \& {Omiya}, M. 2020,
  \aap, 642, A23, \dodoi{10.1051/0004-6361/202038332}

\bibitem[{{Mayor} {et~al.}(2011){Mayor}, {Marmier}, {Lovis}, {Udry},
  {S{\'e}gransan}, {Pepe}, {Benz}, {Bertaux}, {Bouchy}, {Dumusque}, {Lo Curto},
  {Mordasini}, {Queloz}, \& {Santos}}]{2011arXiv1109.2497M}
{Mayor}, M., {Marmier}, M., {Lovis}, C., {et~al.} 2011, arXiv e-prints,
  arXiv:1109.2497.
\newblock \doarXiv{1109.2497}

\bibitem[{{McLaughlin}(1924)}]{1924ApJ....60...22M}
{McLaughlin}, D.~B. 1924, \apj, 60, 22, \dodoi{10.1086/142826}

\bibitem[{{Millholland} \& {Spalding}(2020)}]{2020ApJ...905...71M}
{Millholland}, S.~C., \& {Spalding}, C. 2020, \apj, 905, 71,
  \dodoi{10.3847/1538-4357/abc4e5}

\bibitem[{{Morton}(2015)}]{Morton2015}
{Morton}, T.~D. 2015, {VESPA: False positive probabilities calculator},
  Astrophysics Source Code Library.
\newblock \doeprint{1503.011}

\bibitem[{{Morton} {et~al.}(2016){Morton}, {Bryson}, {Coughlin}, {Rowe},
  {Ravichandran}, {Petigura}, {Haas}, \& {Batalha}}]{Morton2016}
{Morton}, T.~D., {Bryson}, S.~T., {Coughlin}, J.~L., {et~al.} 2016, \apj, 822,
  86, \dodoi{10.3847/0004-637X/822/2/86}

\bibitem[{{Muirhead} {et~al.}(2012){Muirhead}, {Johnson}, {Apps}, {Carter},
  {Morton}, {Fabrycky}, {Pineda}, {Bottom}, {Rojas-Ayala}, {Schlawin},
  {Hamren}, {Covey}, {Crepp}, {Stassun}, {Pepper}, {Hebb}, {Kirby}, {Howard},
  {Isaacson}, {Marcy}, {Levitan}, {Diaz-Santos}, {Armus}, \&
  {Lloyd}}]{2012ApJ...747..144M}
{Muirhead}, P.~S., {Johnson}, J.~A., {Apps}, K., {et~al.} 2012, \apj, 747, 144,
  \dodoi{10.1088/0004-637X/747/2/144}

\bibitem[{{Ohta} {et~al.}(2009){Ohta}, {Taruya}, \&
  {Suto}}]{2009ApJ...690....1O}
{Ohta}, Y., {Taruya}, A., \& {Suto}, Y. 2009, \apj, 690, 1,
  \dodoi{10.1088/0004-637X/690/1/1}

\bibitem[{{Otegi} {et~al.}(2020){Otegi}, {Bouchy}, \&
  {Helled}}]{2020A&A...634A..43O}
{Otegi}, J.~F., {Bouchy}, F., \& {Helled}, R. 2020, \aap, 634, A43,
  \dodoi{10.1051/0004-6361/201936482}

\bibitem[{{Owen} \& {Wu}(2013)}]{2013ApJ...775..105O}
{Owen}, J.~E., \& {Wu}, Y. 2013, \apj, 775, 105,
  \dodoi{10.1088/0004-637X/775/2/105}

\bibitem[{{Owen} \& {Wu}(2017)}]{2017ApJ...847...29O}
---. 2017, \apj, 847, 29, \dodoi{10.3847/1538-4357/aa890a}

\bibitem[{{Rajpaul} {et~al.}(2015){Rajpaul}, {Aigrain}, {Osborne}, {Reece}, \&
  {Roberts}}]{2015MNRAS.452.2269R}
{Rajpaul}, V., {Aigrain}, S., {Osborne}, M.~A., {Reece}, S., \& {Roberts}, S.
  2015, \mnras, 452, 2269, \dodoi{10.1093/mnras/stv1428}

\bibitem[{{Ramirez} \& {Kaltenegger}(2014)}]{2014ApJ...797L..25R}
{Ramirez}, R.~M., \& {Kaltenegger}, L. 2014, \apjl, 797, L25,
  \dodoi{10.1088/2041-8205/797/2/L25}

\bibitem[{{Ricker} {et~al.}(2015){Ricker}, {Winn}, {Vanderspek}, {Latham},
  {Bakos}, {Bean}, {Berta-Thompson}, {Brown}, {Buchhave}, {Butler}, {Butler},
  {Chaplin}, {Charbonneau}, {Christensen-Dalsgaard}, {Clampin}, {Deming},
  {Doty}, {De Lee}, {Dressing}, {Dunham}, {Endl}, {Fressin}, {Ge}, {Henning},
  {Holman}, {Howard}, {Ida}, {Jenkins}, {Jernigan}, {Johnson}, {Kaltenegger},
  {Kawai}, {Kjeldsen}, {Laughlin}, {Levine}, {Lin}, {Lissauer}, {MacQueen},
  {Marcy}, {McCullough}, {Morton}, {Narita}, {Paegert}, {Palle}, {Pepe},
  {Pepper}, {Quirrenbach}, {Rinehart}, {Sasselov}, {Sato}, {Seager},
  {Sozzetti}, {Stassun}, {Sullivan}, {Szentgyorgyi}, {Torres}, {Udry}, \&
  {Villasenor}}]{2015JATIS...1a4003R}
{Ricker}, G.~R., {Winn}, J.~N., {Vanderspek}, R., {et~al.} 2015, Journal of
  Astronomical Telescopes, Instruments, and Systems, 1, 014003,
  \dodoi{10.1117/1.JATIS.1.1.014003}

\bibitem[{{Rossiter}(1924)}]{1924ApJ....60...15R}
{Rossiter}, R.~A. 1924, \apj, 60, 15, \dodoi{10.1086/142825}

\bibitem[{{Rustamkulov} {et~al.}(2022){Rustamkulov}, {Sing}, {Liu}, \&
  {Wang}}]{2022ApJ...928L...7R}
{Rustamkulov}, Z., {Sing}, D.~K., {Liu}, R., \& {Wang}, A. 2022, \apjl, 928,
  L7, \dodoi{10.3847/2041-8213/ac5b6f}

\bibitem[{{Sabotta} {et~al.}(2021){Sabotta}, {Schlecker}, {Chaturvedi},
  {Guenther}, {Mu{\~n}oz Rodr{\'\i}guez}, {Mu{\~n}oz S{\'a}nchez}, {Caballero},
  {Shan}, {Reffert}, {Ribas}, {Reiners}, {Hatzes}, {Amado}, {Klahr}, {Morales},
  {Quirrenbach}, {Henning}, {Dreizler}, {Pall{\'e}}, {Perger}, {Azzaro},
  {Jeffers}, {Kaminski}, {K{\"u}rster}, {Lafarga}, {Montes}, {Passegger}, \&
  {Zechmeister}}]{2021A&A...653A.114S}
{Sabotta}, S., {Schlecker}, M., {Chaturvedi}, P., {et~al.} 2021, \aap, 653,
  A114, \dodoi{10.1051/0004-6361/202140968}

\bibitem[{{Sanchis-Ojeda} {et~al.}(2015){Sanchis-Ojeda}, {Rappaport},
  {Pall{\`e}}, {Delrez}, {DeVore}, {Gandolfi}, {Fukui}, {Ribas}, {Stassun},
  {Albrecht}, {Dai}, {Gaidos}, {Gillon}, {Hirano}, {Holman}, {Howard},
  {Isaacson}, {Jehin}, {Kuzuhara}, {Mann}, {Marcy}, {Miles-P{\'a}ez},
  {Monta{\~n}{\'e}s-Rodr{\'\i}guez}, {Murgas}, {Narita}, {Nowak}, {Onitsuka},
  {Paegert}, {Van Eylen}, {Winn}, \& {Yu}}]{2015ApJ...812..112S}
{Sanchis-Ojeda}, R., {Rappaport}, S., {Pall{\`e}}, E., {et~al.} 2015, \apj,
  812, 112, \dodoi{10.1088/0004-637X/812/2/112}

\bibitem[{{Schlichting} \& {Young}(2022)}]{2022PSJ.....3..127S}
{Schlichting}, H.~E., \& {Young}, E.~D. 2022, \psj, 3, 127,
  \dodoi{10.3847/PSJ/ac68e6}

\bibitem[{{Scott}(2019)}]{Scott2019}
{Scott}, N.~J. 2019, in AAS/Division for Extreme Solar Systems Abstracts,
  Vol.~51, AAS/Division for Extreme Solar Systems Abstracts, 330.15

\bibitem[{{Seifahrt} {et~al.}(2020){Seifahrt}, {Bean}, {St{\"u}rmer}, {Kasper},
  {Gers}, {Schwab}, {Zechmeister}, {Stef{\'a}nsson}, {Montet}, {Dos Santos},
  {Peck}, {White}, \& {Tapia}}]{2020SPIE11447E..1FS}
{Seifahrt}, A., {Bean}, J.~L., {St{\"u}rmer}, J., {et~al.} 2020, in Society of
  Photo-Optical Instrumentation Engineers (SPIE) Conference Series, Vol. 11447,
  Society of Photo-Optical Instrumentation Engineers (SPIE) Conference Series,
  114471F, \dodoi{10.1117/12.2561564}

\bibitem[{{Serrano} {et~al.}(2022){Serrano}, {Gandolfi}, {Hoyer}, {Brandeker},
  {Hooton}, {Sousa}, {Murgas}, {Ciardi}, {Howell}, {Benz}, {Billot},
  {Flor{\'e}n}, {Bekkelien}, {Bonfanti}, {Krenn}, {Mustill}, {Wilson},
  {Osborn}, {Parviainen}, {Heidari}, {Pall{\'e}}, {Fridlund}, {Adibekyan},
  {Fossati}, {Deleuil}, {Knudstrup}, {Collins}, {Lam}, {Grziwa}, {Salmon},
  {Albrecht}, {Alibert}, {Alonso}, {Anglada-Escud{\'e}}, {B{\'a}rczy}, {Barrado
  y Navascues}, {Barros}, {Baumjohann}, {Beck}, {Beck}, {Bieryla}, {Bonfils},
  {Boyd}, {Broeg}, {Cabrera}, {Charnoz}, {Chazelas}, {Christiansen}, {Collier
  Cameron}, {Cort{\'e}s-Zuleta}, {Csizmadia}, {Davies}, {Deline}, {Delrez},
  {Demangeon}, {Demory}, {Dunlavey}, {Ehrenreich}, {Erikson}, {Fortier},
  {Fukui}, {Garai}, {Gillon}, {G{\"u}del}, {H{\'e}brard}, {Heng}, {Huang},
  {Isaak}, {Jenkins}, {Kiss}, {Laskar}, {Latham}, {Lecavelier des Etangs},
  {Lendl}, {Levine}, {Lovis}, {Lund}, {Magrin}, {Maxted}, {Narita},
  {Nascimbeni}, {Olofsson}, {Ottensamer}, {Pagano}, {Pessanha}, {Peter},
  {Piotto}, {Pollacco}, {Queloz}, {Ragazzoni}, {Rando}, {Ratti}, {Rauer},
  {Ribas}, {Ricker}, {Rowden}, {Santos}, {Scandariato}, {Seager},
  {S{\'e}gransan}, {Simon}, {Smith}, {Steller}, {Szab{\'o}}, {Thomas},
  {Twicken}, {Udry}, {Ulmer}, {Van Grootel}, {Vanderspek}, {Viotto}, \&
  {Walton}}]{2022A&A...667A...1S}
{Serrano}, L.~M., {Gandolfi}, D., {Hoyer}, S., {et~al.} 2022, \aap, 667, A1,
  \dodoi{10.1051/0004-6361/202243093}

\bibitem[{{Skrutskie} {et~al.}(2006){Skrutskie}, {Cutri}, {Stiening},
  {Weinberg}, {Schneider}, {Carpenter}, {Beichman}, {Capps}, {Chester},
  {Elias}, {Huchra}, {Liebert}, {Lonsdale}, {Monet}, {Price}, {Seitzer},
  {Jarrett}, {Kirkpatrick}, {Gizis}, {Howard}, {Evans}, {Fowler}, {Fullmer},
  {Hurt}, {Light}, {Kopan}, {Marsh}, {McCallon}, {Tam}, {Van Dyk}, \&
  {Wheelock}}]{2006AJ....131.1163S}
{Skrutskie}, M.~F., {Cutri}, R.~M., {Stiening}, R., {et~al.} 2006, \aj, 131,
  1163, \dodoi{10.1086/498708}

\bibitem[{{Smith} {et~al.}(2012){Smith}, {Stumpe}, {Van Cleve}, {Jenkins},
  {Barclay}, {Fanelli}, {Girouard}, {Kolodziejczak}, {McCauliff}, {Morris}, \&
  {Twicken}}]{2012PASP..124.1000S}
{Smith}, J.~C., {Stumpe}, M.~C., {Van Cleve}, J.~E., {et~al.} 2012, \pasp, 124,
  1000, \dodoi{10.1086/667697}

\bibitem[{{Southworth}(2011)}]{2011MNRAS.417.2166S}
{Southworth}, J. 2011, \mnras, 417, 2166,
  \dodoi{10.1111/j.1365-2966.2011.19399.x}

\bibitem[{{Stassun} {et~al.}(2019){Stassun}, {Oelkers}, {Paegert}, {Torres},
  {Pepper}, {De Lee}, {Collins}, {Latham}, {Muirhead}, {Chittidi},
  {Rojas-Ayala}, {Fleming}, {Rose}, {Tenenbaum}, {Ting}, {Kane}, {Barclay},
  {Bean}, {Brassuer}, {Charbonneau}, {Ge}, {Lissauer}, {Mann}, {McLean},
  {Mullally}, {Narita}, {Plavchan}, {Ricker}, {Sasselov}, {Seager}, {Sharma},
  {Shiao}, {Sozzetti}, {Stello}, {Vanderspek}, {Wallace}, \&
  {Winn}}]{2019AJ....158..138S}
{Stassun}, K.~G., {Oelkers}, R.~J., {Paegert}, M., {et~al.} 2019, \aj, 158,
  138, \dodoi{10.3847/1538-3881/ab3467}

\bibitem[{{Stefansson} {et~al.}(2020){Stefansson}, {Ca{\~n}as}, {Wisniewski},
  {Robertson}, {Mahadevan}, {Maney}, {Kanodia}, {Beard}, {Bender}, {Brunt},
  {Clemens}, {Cochran}, {Diddams}, {Endl}, {Ford}, {Fredrick}, {Halverson},
  {Hearty}, {Hebb}, {Huehnerhoff}, {Jennings}, {Kaplan}, {Levi}, {Lubar},
  {Metcalf}, {Monson}, {Morris}, {Ninan}, {Nitroy}, {Ramsey}, {Roy}, {Schwab},
  {Sigurdsson}, {Terrien}, \& {Wright}}]{2020AJ....159..100S}
{Stefansson}, G., {Ca{\~n}as}, C., {Wisniewski}, J., {et~al.} 2020, \aj, 159,
  100, \dodoi{10.3847/1538-3881/ab5f15}

\bibitem[{{Stumpe} {et~al.}(2014){Stumpe}, {Smith}, {Catanzarite}, {Van Cleve},
  {Jenkins}, {Twicken}, \& {Girouard}}]{2014PASP..126..100S}
{Stumpe}, M.~C., {Smith}, J.~C., {Catanzarite}, J.~H., {et~al.} 2014, \pasp,
  126, 100, \dodoi{10.1086/674989}

\bibitem[{{Stumpe} {et~al.}(2012){Stumpe}, {Smith}, {Van Cleve}, {Twicken},
  {Barclay}, {Fanelli}, {Girouard}, {Jenkins}, {Kolodziejczak}, {McCauliff}, \&
  {Morris}}]{2012PASP..124..985S}
{Stumpe}, M.~C., {Smith}, J.~C., {Van Cleve}, J.~E., {et~al.} 2012, \pasp, 124,
  985, \dodoi{10.1086/667698}

\bibitem[{{Tamura} {et~al.}(2012){Tamura}, {Suto}, {Nishikawa}, {Kotani},
  {Sato}, {Aoki}, {Usuda}, {Kurokawa}, {Kashiwagi}, {Nishiyama}, {Ikeda},
  {Hall}, {Hodapp}, {Hashimoto}, {Morino}, {Inoue}, {Mizuno}, {Washizaki},
  {Tanaka}, {Suzuki}, {Kwon}, {Suenaga}, {Oh}, {Narita}, {Kokubo}, {Hayano},
  {Izumiura}, {Kambe}, {Kudo}, {Kusakabe}, {Ikoma}, {Hori}, {Omiya}, {Genda},
  {Fukui}, {Fujii}, {Guyon}, {Harakawa}, {Hayashi}, {Hidai}, {Hirano},
  {Kuzuhara}, {Machida}, {Matsuo}, {Nagata}, {Ohnuki}, {Ogihara}, {Oshino},
  {Suzuki}, {Takami}, {Takato}, {Takahashi}, {Tachinami}, \&
  {Terada}}]{2012SPIE.8446E..1TT}
{Tamura}, M., {Suto}, H., {Nishikawa}, J., {et~al.} 2012, in \procspie, Vol.
  8446, Ground-based and Airborne Instrumentation for Astronomy IV, 84461T,
  \dodoi{10.1117/12.925885}

\bibitem[{{Van Eylen} {et~al.}(2019){Van Eylen}, {Albrecht}, {Huang},
  {MacDonald}, {Dawson}, {Cai}, {Foreman-Mackey}, {Lundkvist}, {Silva Aguirre},
  {Snellen}, \& {Winn}}]{2019AJ....157...61V}
{Van Eylen}, V., {Albrecht}, S., {Huang}, X., {et~al.} 2019, \aj, 157, 61,
  \dodoi{10.3847/1538-3881/aaf22f}

\bibitem[{{Van Eylen} {et~al.}(2021){Van Eylen}, {Astudillo-Defru}, {Bonfils},
  {Livingston}, {Hirano}, {Luque}, {Lam}, {Justesen}, {Winn}, {Gandolfi},
  {Nowak}, {Palle}, {Albrecht}, {Dai}, {Campos Estrada}, {Owen},
  {Foreman-Mackey}, {Fridlund}, {Korth}, {Mathur}, {Forveille}, {Mikal-Evans},
  {Osborne}, {Ho}, {Almenara}, {Artigau}, {Barrag{\'a}n}, {Barros}, {Bouchy},
  {Cabrera}, {Caldwell}, {Charbonneau}, {Chaturvedi}, {Cochran}, {Csizmadia},
  {Damasso}, {Delfosse}, {De Medeiros}, {D{\'\i}az}, {Doyon}, {Esposito},
  {F{\H{u}}r{\'e}sz}, {Figueira}, {Georgieva}, {Goffo}, {Grziwa}, {Guenther},
  {Hatzes}, {Jenkins}, {Kabath}, {Knudstrup}, {Latham}, {Lavie}, {Lovis},
  {Mennickent}, {Mullally}, {Murgas}, {Narita}, {Pepe}, {Persson}, {Redfield},
  {Ricker}, {Santos}, {Seager}, {Serrano}, {Smith}, {Su{\'a}rez Mascare{\~n}o},
  {Subjak}, {Twicken}, {Udry}, {Vanderspek}, \& {Zapatero
  Osorio}}]{2021MNRAS.507.2154V}
{Van Eylen}, V., {Astudillo-Defru}, N., {Bonfils}, X., {et~al.} 2021, \mnras,
  507, 2154, \dodoi{10.1093/mnras/stab2143}

\bibitem[{{Vanderburg} \& {Johnson}(2014{\natexlab{a}})}]{Vanderburg}
{Vanderburg}, A., \& {Johnson}, J.~A. 2014{\natexlab{a}}, \pasp, 126, 948,
  \dodoi{10.1086/678764}

\bibitem[{{Vanderburg} \& {Johnson}(2014{\natexlab{b}})}]{2014PASP..126..948V}
---. 2014{\natexlab{b}}, \pasp, 126, 948, \dodoi{10.1086/678764}

\bibitem[{{Weiss} {et~al.}(2018){Weiss}, {Marcy}, {Petigura}, {Fulton},
  {Howard}, {Winn}, {Isaacson}, {Morton}, {Hirsch}, {Sinukoff}, {Cumming},
  {Hebb}, \& {Cargile}}]{2018AJ....155...48W}
{Weiss}, L.~M., {Marcy}, G.~W., {Petigura}, E.~A., {et~al.} 2018, \aj, 155, 48,
  \dodoi{10.3847/1538-3881/aa9ff6}

\bibitem[{{Winn} {et~al.}(2018){Winn}, {Sanchis-Ojeda}, \&
  {Rappaport}}]{2018NewAR..83...37W}
{Winn}, J.~N., {Sanchis-Ojeda}, R., \& {Rappaport}, S. 2018, \nar, 83, 37,
  \dodoi{10.1016/j.newar.2019.03.006}

\bibitem[{{Zechmeister} \& {K{\"u}rster}(2009)}]{2009A&A...496..577Z}
{Zechmeister}, M., \& {K{\"u}rster}, M. 2009, \aap, 496, 577,
  \dodoi{10.1051/0004-6361:200811296}

\bibitem[{{Zeng} {et~al.}(2016){Zeng}, {Sasselov}, \&
  {Jacobsen}}]{2016ApJ...819..127Z}
{Zeng}, L., {Sasselov}, D.~D., \& {Jacobsen}, S.~B. 2016, \apj, 819, 127,
  \dodoi{10.3847/0004-637X/819/2/127}

\bibitem[{{Zeng} {et~al.}(2019){Zeng}, {Jacobsen}, {Sasselov}, {Petaev},
  {Vanderburg}, {Lopez-Morales}, {Perez-Mercader}, {Mattsson}, {Li}, {Heising},
  {Bonomo}, {Damasso}, {Berger}, {Cao}, {Levi}, \&
  {Wordsworth}}]{2019PNAS..116.9723Z}
{Zeng}, L., {Jacobsen}, S.~B., {Sasselov}, D.~D., {et~al.} 2019, Proceedings of
  the National Academy of Science, 116, 9723, \dodoi{10.1073/pnas.1812905116}

\end{thebibliography}
\bibliographystyle{aasjournal}



\end{document}